\documentclass[preprint]{aastex631}
\usepackage[T1]{fontenc}

\def\simgr{\,\hbox{\hbox{$ > $}\kern -0.8em \lower 1.0ex\hbox{$\sim$}}\,}
\def\simle{\,\hbox{\hbox{$ < $}\kern -0.8em \lower 1.0ex\hbox{$\sim$}}\,}

\newcommand{\nswift}{Neil Gehrels Swift Observatory}
\newcommand{\swift}{Swift}
\newcommand{\fermi}{Fermi}
\newcommand{\xmm}{XMM-Newton}

\newcommand{\mspone}{PSR~J1628$-$3205}
\newcommand{\msptwo}{PSR~J1048$+$2339}
\newcommand{\mspthree}{XMMU J083850.38$-$282756.8}
\newcommand{\fgl}{3FGL~J0838.8$-$2829}
\newcommand{\obj}{1FGL~J0523.5$-$2529}

\shortauthors{Halpern et al.}
\shorttitle{Flaring of \obj}

\begin{document}
\title{Luminous Optical and X-ray Flaring of the Putative Redback Millisecond Pulsar \obj}

\author[0000-0003-4814-2377]{Jules P. Halpern}
\author[0000-0002-6341-4548]{Karen I. Perez}
\author[0000-0002-9870-2742]{Slavko Bogdanov}
\affiliation{Department of Astronomy and Columbia Astrophysics Laboratory, Columbia University, 550 West 120th Street, New York, NY 10027-6601, USA; jph1@columbia.edu}

\begin{abstract}
Several redback and black widow millisecond pulsar binaries have episodes of flaring in X-rays and optical.  We initially detected such behavior from the \fermi\ selected redback candidate \obj\ during optical time-series monitoring.  Triggered observations with the \nswift\ over the next $\approx100$ days showed episodic flaring in X-rays with luminosity up to $8\times10^{33}$ erg~s$^{-1}$ ($\sim100$ times the minimum), and a comparable luminosity in the optical/UV, with similar power-law spectra of $f_{\nu}\propto\nu^{-0.7}$. These are the most luminous flares seen in any non-accreting ``spider'' pulsar system, which may be related to the large size of the companion through the fraction of the pulsar wind that it or its ablated wind intercepts.  Simultaneously with an optical flare, we see Balmer-line and \ion{He}{1} emission, not previously known in this object, which is evidence of a stellar wind that may also inhibit detection of radio pulsations.  The quiescent optical light curves, while dominated by ellipsoidal modulation, show evidence of variable non-uniform temperature that could be due either to large starspots or asymmetric heating of the companion by the pulsar.  This may explain a previous measurement of unusual non-zero orbital eccentricity as, alternatively, distortion of the radial-velocity curve by the surface temperature distribution of the large companion.
\end{abstract}


\section{Introduction\label{sec:intro}}

Millisecond pulsars (MSPs) have shorter spin periods and smaller spin-down rates than ordinary pulsars, which in the dipole spindown paradigm implies weaker surface magnetic fields ($B_s\sim10^{8-9}$~G) and much older ages ($\sim10^9$~yr).  Unlike canonical pulsars, the majority of MSPs have binary companions, either white dwarfs or low-mass, evolved stars.  The standard model for their origin is spin-up (recycling) of the neutron star by accretion during a low-mass X-ray binary (LMXB) stage \citep{alp82,rad82}, with the radio pulsar turning on after accretion stops.  Accretion-induced collapse of a white dwarf has been considered an alternative formation scenario \citep{fer07,tau13}. 

Of 3329 cataloged pulsars\footnote{https://www.atnf.csiro.au/research/pulsar/psrcat/ (version 1.67)} \citep{man05}, 501 have spin periods $<20$~ms and $B_s<10^{10}$~G, a combination that reliably indicates a recycled origin.  Low-$B$-field pulsars with slightly longer periods may also have been recycled, albeit ``mildly''.  There are also 20 accretion-powered MSPs known as of 2020 \citep{pap22}, all in X-ray transient systems with orbital periods $<1$~day, that may eventually become radio MSPs.

The Large Area Telescope on the Fermi Gamma-ray Space Telescope detects many known and new MSPs\footnote{https://confluence.slac.stanford.edu/display/GLAMCOG/Public+List+of+LAT-Detected+Gamma-Ray+Pulsars}.  Of the 276 $\gamma$-ray pulsars detected to date, 127 are MSPs, of which 95 are Fermi discoveries, e.g., \citet{bat21}. Follow-up radio pulsar searches of unidentified Fermi source positions identifies the pulsar counterpart, and continued timing measures its orbital parameters.  Discovery of the corresponding $\gamma$-ray pulsations often follows.  Direct pulsar searches of the $\gamma$-rays themselves have also revealed MSPs \citep{ple12,cla18,cla21,nie20b}; in two of these cases the pulsar remains undetected in radio.

An important Fermi discovery is the abundance of black widows and redbacks, previously rare MSPs in which the relativistic pulsar wind heats the photosphere of a companion and helps drive a wind. With orbital periods generally $\simle 1$~day, black widows have degenerate companions of $<0.05\,M_{\odot}$, while redbacks have non-degenerate $0.1 - 1\,M_{\odot}$ companions that are generally hotter and of lower density than main-sequence stars of the same mass.  These distinct sub-classes were recognized \citep{rob11,rob13} as a bimodal mass distribution became apparent among the Fermi MSP identifications.  \citet{hui19} catalogued 44 black widows (27 in the Galactic field and 17 in globular clusters), and 26 redbacks (14 in the Galactic field and 12 in globular clusters). 

Redback companions are close to filling their Roche lobes \citep{str19}, and their winds often obscure the radio pulsar signal for a large fraction of the orbit (e.g., \citealt{cam16,den16,cla21,cor21}). The winds are sometimes seen in optical emission lines \citep{hal17b,str19}, while nonthermal emission, likely due to shocks from the collision of the pulsar wind and the stellar wind, is ubiquitous in X-rays \citep{bog05,bog11,bog14,aln18}.  When good photon statistics are available, the X-ray flux is often seen to be modulated around the orbit, which has been interpreted in terms of the geometry of the intrabinary shock and relativistic beaming of synchrotron radiation by the emitting particles \citep{rom16,wad17,cor22}.

The photosphere of the companion can be heated either by high-energy photons from the shock region, or direct bombardment by pulsar wind particles \citep{san17}.  The resulting optical modulation can be dominated by the contrast between the heated and the ``dark'' side of the companion \citep{bre13,sch14}, or, if the heating effect is small, by ellipsoidal modulation of the light from the tidally distorted star \citep{li14,bel16,van16,sha17}, with large starspots possibly playing a role.

All of these effects can also vary on timescales of minutes to months. A change in photospheric heating sometimes occurs over weeks and months, while minutes long flares increase the luminosity by a factor of 10 or more in X-rays, and up to $\sim1$ mag in optical, as seen in \msptwo\ \citep{cho18} and \fgl\ \citep{hal17a}.  In the latter redback candidate, a flare was seen simultaneously in X-ray and optical by \xmm. One benefit of these signature effects is that putative redbacks are easily recognized even if their millisecond pulsations have not yet been detected. The detailed properties of well-studied redbacks were summarized by \citet{str19}, including six of the putative ones. More recent identifications of candidate redbacks among Fermi sources are
4FGL~J2333.1$-$5527 \citep{swi20},
4FGL~J0935.3+0901 \citep{wan20},
4FGL~J0940.3$-$7610 \citep{swi21},
and 4FGL~J1702.7$-$5655 \citep{corb22}.
Pulsations have not been detected from these last four despite dedicated searches in radio \citep{cam15,cam16,zhe22}.  Of these, 4FGL~J0935.3+0901 has both X-ray flares \citep{zhe22} and optical flares \citep{hal22}, although it is not clear if it is a redback or a black widow.

Some relatives of redbacks have an accretion disk, but one that is truncated at the inner boundary by the rotating pulsar magnetosphere or pulsar wind. They are easily recognized by their pattern of temporal variability, a distinctive moding behavior in the X-ray and optical that is unique to this class of X-ray binary \citep{bog15a,bog15b,dem15}.  [Although, similar behavior has been observed in several cataclysmic variables \citep{sca17,sca22,lit22,duf22} and attributed to a similar mechanism, magnetic gating of the accretion.]  Known as transitional millisecond pulsars (tMSPs) in the sub-luminous disk state, the majority of these were also identified from Fermi sources.

tMSPs may be the evolutionary link between the MSPs and their LMXB progenitors \citep{arc09}. Three tMSPs have been observed to transition between the disk state and the redback pulsar state, PSR~J1023+0038 \citep{sta14}, XSS~J12270$-$4859 \citep{bas14}, and PSR~J1824$-$2452I in the globular cluster M28 \citep{pap13}.  One of these, PSR~J1824$-$2452I, transitioned directly between LMXB and redback states.  There are an additional five putative tMSPs identified whose spin periods are not yet known because they have not yet been observed in a pulsar state \citep{bog15b,str16,mil20,cot21,str21}. See \citet{pap22} for a review of tMSPs, their accretion physics, and their relation to redbacks.

Evolutionary pathways that lead to the transformation of LMXBs into black widows and redbacks have been modeled using physical processes that are likely relevant, although difficult to quantify.  These effects include irradiation feedback by accretion luminosity during Roche-lobe overflow, evaporation of the companion by pulsar wind and its radiation, and loss of orbital angular momentum by magnetic braking \citep{ben12,ben14,ben15,che13,dev20,gin20,gin21,jia15,jia16}.  The resulting evolutionary tracks are tested against the observed orbital periods, companion masses, and Roche-lobe filling factors where measured.  Accretion-induced collapse of a white dwarf to a MSP has also been considered as a formation mechanism for redbacks and black widows \citep{liu17,abl19}.

These studies generally found that stronger evaporation of the companion leads to the formation of redbacks, which may go through repeated cycles of accretion and shrinkage within the Roche-lobe.  Weaker evaporation leads to black widows, but evaporation alone is not strong enough to completely destroy the companion.  Black widows require a mechanism of angular momentum loss to remove matter by Roche-lobe overflow.  Theoretically, both redback and black widow companions should not be much smaller than their Roche lobes, which is broadly in accord with the interpretation of observations \citep{str19,dra19}.

The subject of this paper is the first one of the putative redbacks to be identified without a spin period, \obj.  It is less studied than most of the others.  \citet{str14} identified its X-ray counterpart with the \nswift, and discovered its 0.688~day optical period in the Catalina Real-Time Transient Survey (CRTS; \citealt{dra09}). Radial velocity spectroscopy of the late G or early K type star, together with a measurement of its rotational broadening, indicated a mass ratio of $0.61\pm0.06$, and a companion mass in the range $0.8-1.3\,M_{\odot}$ for an assumed NS mass in the range $1.4-2.0\,M_{\odot}$.   These parameters require an orbital inclination in the range $57^{\circ}<i<80^{\circ}$.

The companion mass of \obj\ is the largest among redbacks \citep{str19}.  In addition, an orbital eccentricity of $0.040\pm0.006$ was measured, which is unexpected for such old systems.  Other redback eccentricities are consistent with zero.  The evidence for a pulsar in the system was less direct than in most other such candidates.  The optical light curve in CRTS is consistent with ellipsoidal modulation and no heating.  There was no evidence for emission lines in the spectra, and no flaring in X-rays or optical, although very little if any monitoring had been done.  

In this paper, we present optical time-series observations, spectra, and Swift monitoring that reveal multiple epochs of flaring behavior and variable heating in \obj, confirming that a relatively energetic MSP must be present even though its pulsations have not yet been detected.  We assume a distance of 2.24~kpc from the Gaia measured parallax of $0.446\pm0.042$~mas in its extended third data release (EDR3; \citealt{bro21}).  Section~\ref{sec:obs} describes the optical data and results from MDM Observatory,  while Section~\ref{sec:tess} examines light curves from the Transiting Exoplanet Survey Satellite (TESS; \citealt{ric15}).  Section~\ref{sec:surveys} compiles long-term survey results from the CRTS, the All-Sky Automated Survey for Supernovae (ASAS-SN; \citealt{jay19}), and the Zwicky Transient Facility (ZTF; \citealt{bel19}).  Observations from the \swift\ X-ray Telescope (XRT) and UV/Optical Telescope (UVOT) are presented in Sections~\ref{sec:xrt} and~\ref{sec:uvot}, respectively.  Interpretation of the results follows in Section~\ref{sec:disc}.  The main conclusions are summarized in Section~\ref{sec:conc}.

\section{MDM Optical Observations\label{sec:obs}}

\begin{deluxetable}{lccc}
\label{tab:optlog}
\tablecolumns{4} 
\tablewidth{0pt} 
\tablecaption{Log of $R$-band Time-Series Photometry}
\tablehead{
\colhead{Date} & \colhead{Exposure\tablenotemark{a}} &
\colhead{Time} & \colhead{Phase} \\
\colhead{(UT)} & \colhead{(s)} & \colhead{(UTC)} &
\colhead{($\phi$)}
}
\startdata
2014 Dec 23  & $3\times15$  & 05:07--07:39 & 0.905--0.057 \\
2014 Dec 23  & $3\times15$  & 04:55--07:33 & 0.346--0.506 \\
2014 Dec 27  & $3\times15$  & 03:44--07:30 & 0.636--0.861 \\
2014 Dec 28  & $3\times15$  & 03:42--07:29 & 0.087--0.313 \\
2014 Dec 29  & $3\times15$  & 03:40--06:56 & 0.536--0.733 \\
2020 Jan 23  & 30           & 02:03--03:22 & 0.321--0.400 \\
2020 Jan 24  & 30           & 01:45--04:55 & 0.755--0.947 \\
2020 Jan 25  & 30           & 02:01--07:50 & 0.225--0.576 \\
2020 Jan 26  & 30           & 01:41--07:40 & 0.658--0.020 \\
2020 Feb 6   & 30           & 01:49--07:03 & 0.650--0.967 \\
2020 Feb 7   & 30           & 01:48--07:00 & 0.102--0.417 \\
2020 Nov 12  & 30           & 06:19--12:37 & 0.822--0.176 \\
2020 Nov 13  & 30           & 06:05--12:10 & 0.261--0.612 \\
2020 Nov 14  & 30           & 06:07--12:29 & 0.716--0.102 \\
2020 Nov 15  & 30           & 05:58--12:31 & 0.160--0.557 \\
2020 Nov 16  & 30           & 05:49--12:27 & 0.605--0.007 \\
2020 Dec 17  & 60           & 03:49--10:20 & 0.534--0.928 \\
2021 Jan 7   & 60           & 02:44--08:56 & 0.071--0.360 \\
2021 Jan 8   & 60           & 02:25--08:55 & 0.418--0.813 \\
2021 Jan 9   & 60           & 03:01--08:55 & 0.908--0.265 \\
2021 Jan 10  & 60           & 02:17--08:50 & 0.317--0.714 \\
2021 Jan 11  & 60           & 02:09--08:47 & 0.762--0.163 \\
2021 Nov 12  & 60           & 06:08--12:46 & 0.231--0.632 \\
2021 Nov 13  & 60           & 06:03--12:42 & 0.679--0.083 \\
2021 Nov 14  & 60           & 05:59--12:38 & 0.128--0.531 \\
2021 Nov 15  & 60           & 05:55--12:34 & 0.578--0.980 \\
\enddata
\tablenotetext{a}{Points in Figure~\ref{fig:time-series}a for 2014 December are the average of three 15~s exposures.}
\end{deluxetable}

We used the MDM Observatory 1.3~m McGraw-Hill telescope for time-series photometry
of \obj\ in the $R$-band.  Thinned, backside-illuminated CCDs were
windowed to achieve an efficient read/prep cycle time of 3~s compared with
the exposure times of 15, 30, or 60~s as listed in the observing log
(Table~\ref{tab:optlog}). 
To create the light curves, differential photometry was performed with respect to
a nearby comparison star.  Two methods were used to calibrate the comparison star.
First, we measured its $R$ magnitude using \citet{lan92} standard stars on a photometric
night.  Second, we transformed its Pan-STARRS $i$ and $r$ magnitudes to $R$.  The results 
agree to 0.01~mag.   We also obtained two spectra on the MDM~2.4~m Hiltner
telescope using OSMOS, the Ohio State Multi-Object Spectrograph \citep{mar11}.

\begin{figure}
  \centerline{
\includegraphics[angle=0.,width=1.05\linewidth]{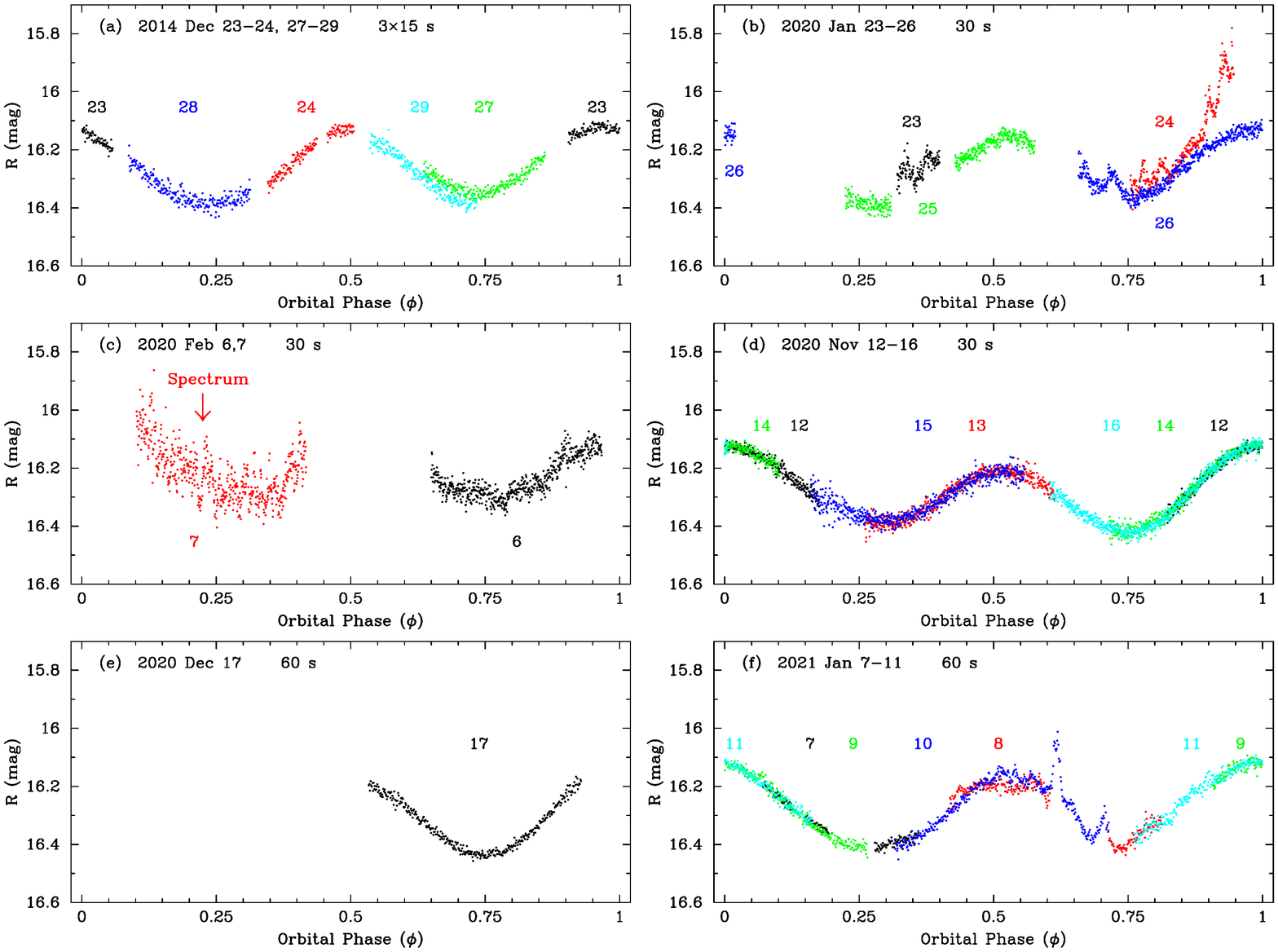}
}
\vspace{-0.5in}
\centerline{
\includegraphics[angle=0.,width=1.05\linewidth]{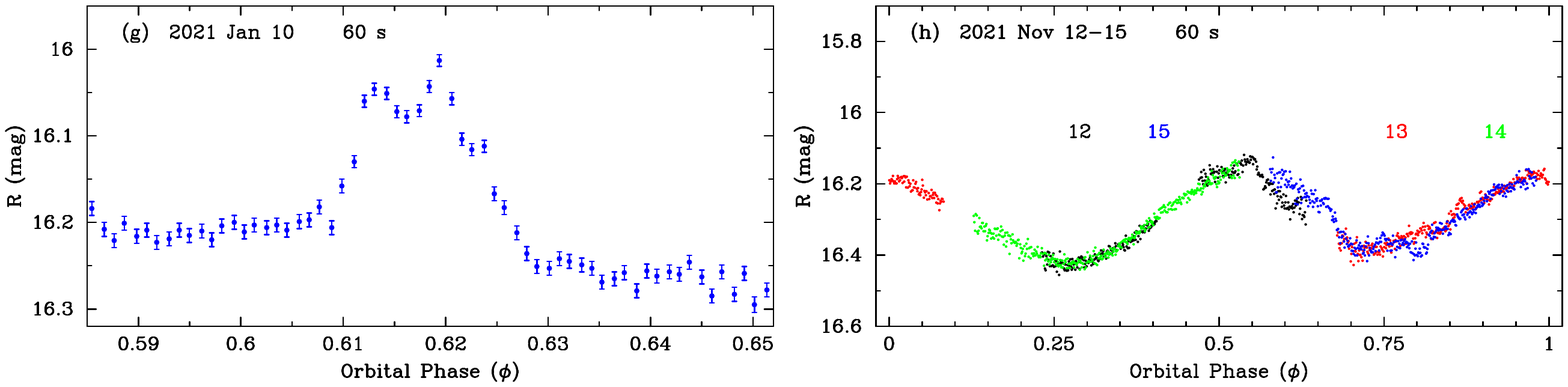}
}
\vspace{-3.8in}
  \caption{$R$-band time-series photometry of \obj\ on the MDM 1.3~m,
    folded according to the orbital ephemeris of \cite{str14}.
    A log of the observations is given in Table~\ref{tab:optlog}.
    Dates and exposure times per point are shown in each panel.  The
    the data are color coded and labeled by the day of the month.
    A spectrum (shown in Figure~\ref{fig:spectra}) was obtained
    during the rapidly flaring state on 2020 February~7 in (panel c).
    Panel g is an expanded view of the largest flare in panel f.
}
\label{fig:time-series}
\end{figure}

Figure~\ref{fig:time-series} shows our 26 nights of photometry grouped into
the seven observing runs: one in 2014, and six in 2020--2021.  Given the
16.5~hr orbital period and southerly declination of \obj, it is possible
to cover {\it at most\/} 0.4 orbit cycles per night, while taking several
nights to (almost) sample the full range of phase.  The light curves are
color coded by night in Figure~\ref{fig:time-series}, grouped by observing
run and graphed as a function of orbital phase.

All computations of orbital phase in this paper use the spectroscopic ephemeris
of \citet{str14}: $P=0.688134\pm0.000028$~d, $T_{0.75}=2456577.64636\pm0.0037$~(BJD), and Gaia-CRF3 position
R.A.=$05^{\rm h}23^{\rm m}16.\!^{\rm s}931$, decl.=$-25^{\circ}27^{\prime}37.\!^{\prime\prime}13$ (\citealt{bro21}), which is referenced for proper motion to
epoch 2016.0.  Here $T_{0.75}$ is the epoch of superior conjunction of the companion star,
which is phase $\phi=0.75$, while $\phi=0$ is the ascending node of the pulsar according to the radio pulsar convention.
The uncertainty in the period measured in 2013--2014 extrapolates to an expected
error in orbital phase of 0.15 by the beginning of 2021, but our light curves
empirically suggest that the extrapolation is more precise than predicted.  However, the light curves
are seen to depart in character from a pure ellipsoidal model in ways that will be described below.  Therefore, it is not obvious from these light curves how to generate a new photometric ephemeris that could be used to verify or refine the old spectroscopic ephemeris.  The extrapolation of the old ephemeris therefore remains a source of uncertainty in absolute phasing.

We next describe the basic features of the data by individual observing runs:
\newpage
\subsection{2014 December 23--29}

Our initial run was designed to obtain a more precise light curve than 
was previously available only from the CRTS \citep{str14}.
Figure~\ref{fig:time-series}a shows the result, which was largely as expected,
a light curve dominated by the effect of ellipsoidal modulation of a tidally
distorted star.  The quadrature points at $\phi=0$ and 0.5 are maxima of
equal brightness, as expected.  The minima, however are unusual in that
superior conjunction of the companion ($\phi=0.75$) is slightly brighter
than inferior conjunction, the opposite of what is expected for a Roche-lobe
filling star, in which the L1 point of the star is the dimmest due to
gravity darkening and limb darkening.  This might indicate that there
is in fact some pulsar heating of the ``day'' side of the companion,
which brightens the superior conjunction.

Furthermore, it is apparent that at least one of the nights that sampled
around superior conjunction is discrepant.  At first we assumed that this
was a calibration problem, which could be fixed by raising the
December~29 data by $\approx0.035$~mag.  But we cannot identify a specific
cause of this discrepancy, and our subsequent observations of variability
now lead us to believe that the change could be real.

\subsection{2020 January 23--26}

We returned to \obj\ in 2020 January in order to add it to our program to characterize the light curves of redbacks and their variability.  From the first exploratory observation on January~23,
new behavior consisting of rapid variability was apparent.  A high
point of $R=15.8$ (above the quiescent magnitude of 16.15) was reached on January~24, when the star was continuing
to brighten with multiple, overlapping flares through the end of the series
(red points in Figure~\ref{fig:time-series}b). It is possible that the peak of the flaring was missed. Based on these results we
triggered a series of \swift\ observations, which began on January 26 and
detected X-ray flaring for several days thereafter (see Section~\ref{sec:xrt}).
Intermittent flaring continued to be observed at MDM through the end of this run on January~26.

\begin{figure}
  \centerline{
\includegraphics[angle=0.,width=1.\linewidth]{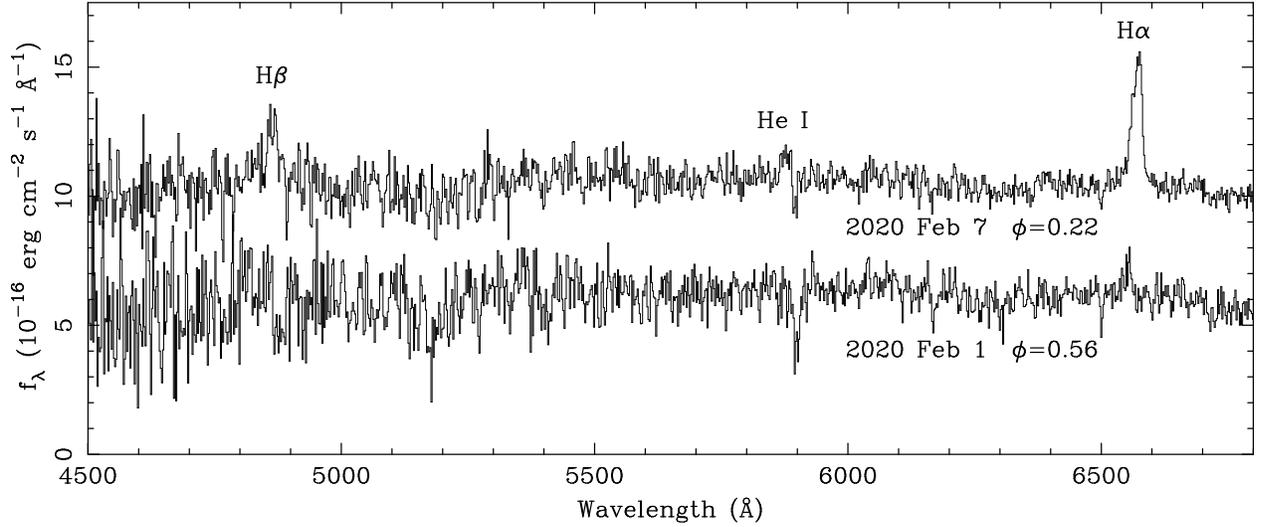}
  }
\vspace{-2.4in}
  \caption{Two 15 minute spectra of \obj\ obtained with the MDM 2.4-m in
    2020 February.  Their fluxes have been offset for clarity.
    Orbital phases are indicated.  Balmer and \ion{He}{1}~$\lambda5876$
    emission lines are highly variable.  The spectrum on February 7
    was simultaneous with time-series photometry in
    Figure~\ref{fig:time-series}c, when \obj\ was in a rapidly flaring state and the emission lines were stronger.
}
\label{fig:spectra}
\end{figure}

\begin{figure}
  \centerline{
\includegraphics[angle=0.,width=1.\linewidth]{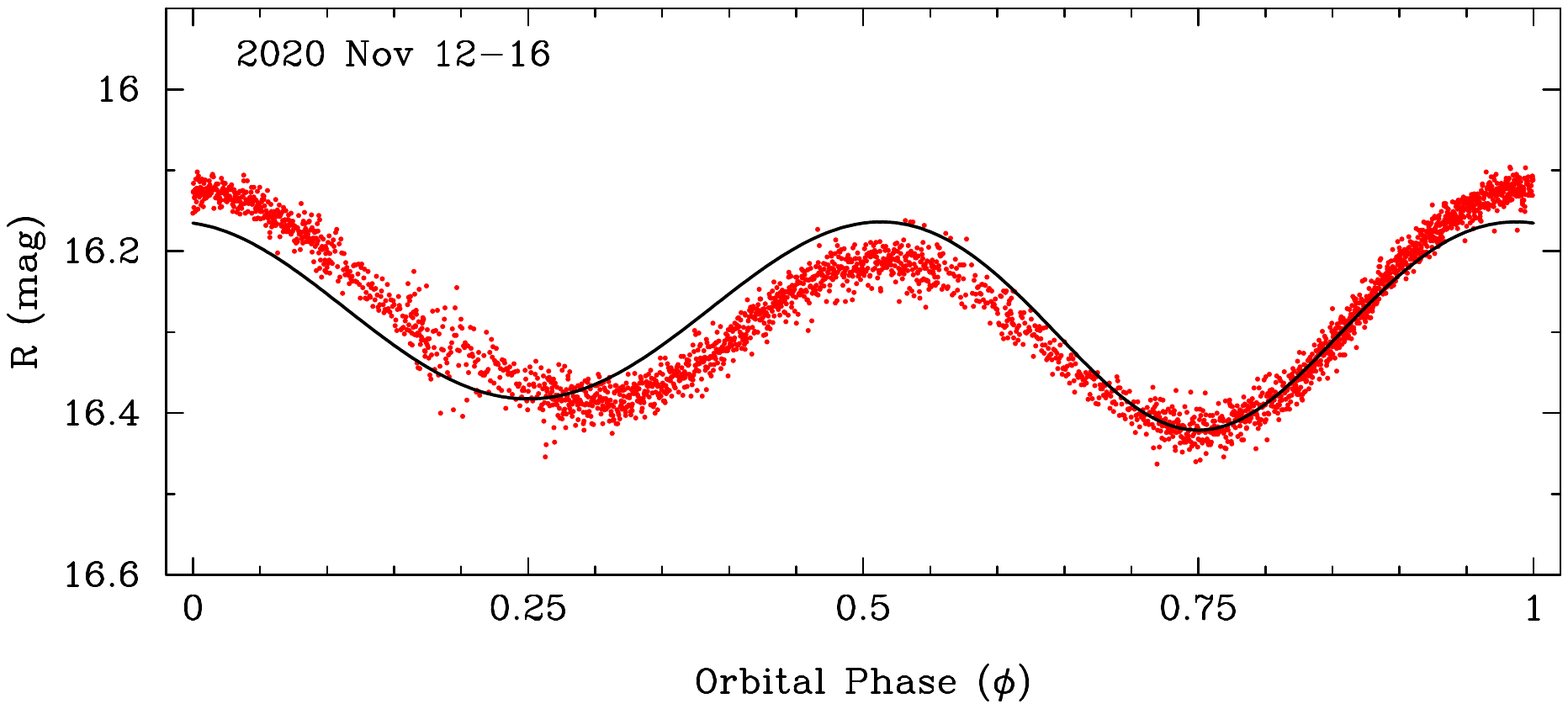}
  }
\vspace{-6.1in}
  \caption{A model of ellipsoidal modulation \citep{mor93,gom21} is superposed
  on the data from Figure~\ref{fig:time-series}d. The orbital inclination $i=67^{\circ}$ and mass ratio $q=0.61$ are from the spectroscopic study of \citet{str14}.  A roche-lobe filling factor $f=0.99$, linear limb-darkening coefficient 0.65, and gravity-darkening coefficient 0.45 ({\it not\/} the gravity-darkening exponent; see \citealt{tor21}) are assumed.  This is not a fit, but an illustration of the inadequacy of any purely ellipsoidal model. See Section~\ref{sec:disc_a} for discussion.
}
\label{fig:model}
\end{figure}

\subsection{2020 February 1,6,7}

We obtained an optical spectrum of \obj\ on 2020 February~1 and found a late-type stellar spectrum (Figure~\ref{fig:spectra}), similar to those of \citet{str14}, but with weak H$\alpha$ emission as a new feature.  On February~6 and 7, time-series photometry revealed a new type of continuous, rapid flaring behavior (Figure~\ref{fig:time-series}c).  This is real variability, not noisy statistics.  The flares peak $\approx0.3-0.4$~mag above the quiescent level, similar to the smoother, more isolated flares seen on other nights.  But the rapid temporal behavior is qualitatively different.  Although the February~7 data may give an impression of possibly being periodic, a power spectrum does not pick out a coherent or quasi-periodic signal.  We also obtained another spectrum simultaneous with the time-series on February~7, which shows much stronger Balmer emission lines as well as \ion{He}{1}~$\lambda5876$, the brightest expected \ion{He}{1} line in this spectral range.

The equivalent width of H$\alpha$ increased from $\approx2.8$~\AA\ on February~1 to $\approx18$~\AA\ (or larger, depending on the extent of its broad wing) on February 7.  In addition, the velocities of the line peaks differ by $\approx1000$~km~s$^{-1}$, while the FWHM on February~7 is also $\approx1000$~km~s$^{-1}$.  The line broadening and wavelength shift of H$\alpha$ evident between the two spectra must be contributed to by velocity structure of the ablated stellar wind, not just orbital dynamics, for which \citet{str14} found a radial velocity amplitude of only $190$~km~s$^{-1}$.  In view of the extreme variability seen simultaneously on February~7, and the lack of emission lines in the \citet{str14} spectra, it seems reasonable to conclude that emission lines are connected with flaring behavior, which was probably not occurring in late 2013 -- early 2014. 

\subsection{2020 November 12--16}

Five nights of time-series photometry in 2020 November did not detect any flaring. There are episodes of noisier than usual data in Figure~\ref{fig:time-series}d, but these are due to poor seeing and/or thin clouds, especially at the high-airmass ends of the series.  However, it is now easier to see that there are subtle differences in the character of this quiescent light curve from the one in 2014 December, which is also largely quiescent.  We observe three types of changes:

First, the maxima at $\phi=0$ and 0.5 are no longer equal, with the one at $\phi=0$ being brighter by $\approx0.06$~mag, a significant fraction of the total modulation of $0.3$~mag.  The same effect
was observed by \citet{pal20} in the light curve of \obj\ as observed by TESS in 2018 November --  2019 January.  In the absence of evidence for pulsar heating, which would affect $\phi=0.75$ the most, it is difficult to appeal to asymmetric heating of the companion as the cause of the mismatched peaks.
A possible explanation is a large starspot or spots that preferentially darken one side of the star, which would be the orbit-trailing side to make it darker at $\phi=0.5$.  We will revisit this idea of starspots in the context of how they might affect the radial velocity curve (Section~\ref{sec:disc_b}).

The second change is in the minimum at $\phi=0.75$.  It is now lower than the one at $\phi=0.25$, reversing the trend from 2014 December in both relative and absolute sense.  That is, the brightness at $\phi=0.25$ remained the same, while $\phi=0.75$ became dimmer.  This could be attributed to a decrease in heating by the pulsar, either directly, or indirectly through the intermediary of an intrabinary shock.

The third change is a shift in the phase of the first minimum (inferior conjunction of the companion) from $\phi=0.25$ to $\approx0.30$.  The two minima are no longer exactly $0.50$ apart in phase.  Note that at $\phi=0.25$ we are viewing the ``dark'' side of the star, the side least likely to be affected by pulsar heating.  Together with the unequal maxima, this delay of the minimum could indicate that the trailing side of the star is darker than the leading side.

While the 2000 November data are the simplest, being free of flares and other variability, they cannot be modelled entirely by ellipsoidal modulation.  The deviations are illustrated in Figure~\ref{fig:model}, where we have superposed an analytic ellipsoidal model \citep{mor93,gom21} incorporating the spectroscopic binary parameters from \citet{str14}.  Similar difficulties would attend the fitting of any of the data in Figure~\ref{fig:time-series}, and are discussed in more detail in Section~\ref{sec:disc_a}.

\subsection{2020 December 17}

Only one night in 2020 December was used for \obj, resulting in coverage only of the minimum at $\phi=0.75$ (Figure~\ref{fig:time-series}e).  It is clear from this light curve that the minimum is as low as, or even slight lower than a month earlier. So there is no evidence for pulsar heating.

\subsection{2021 January 7--11}

Renewed flaring was detected on two of the five nights observed in 2021 January (Figure~\ref{fig:time-series}f).  These share many of the features of previous results, with the example of a strong and well-resolved flare on January 10 that lasts $\approx25$~minutes at $\phi=0.62$. An expanded view of this flare is shown in Figure~\ref{fig:time-series}g.  The light curves create an illusion that flares on January 8 and 10 may be repeating at the same phases, 0.62 and 0.71, but there is no evidence for this since most of the overlapping January 8 run (red points) is missing due to clouds.  The minimum at $\phi=0.25$ still appears to be late.

\subsection{2021 November 12--15}

Low-level flaring was again detected.  A failure of the telescope drive interrupted the time-series for 1~hour on November 12 (black points in Figure~\ref{fig:time-series}h). The minimum at $\phi=0.25$ still appears to be late.
  
\section{TESS Observation (2018 November 15 - 2019 January 6)\label{sec:tess}}
  
Light curves with 120~s cadence were obtained by TESS spanning 52 days in 2018--2019, with three gaps totaling $\approx7$~days.  These were first described by \citet{pal20}, where we see that the period listed differs from the one we are using by an amount that is certainly incompatible.  Actually, two orbital period values are listed, 0.68505(62)~days and 0.68973(17)~days, deriving respectively from the fundamental and first harmonic in the power spectrum.  The harmonic is more precise because it contains most of the power from the double-peaked light curve, $\approx94\%$ by our own calculation.  But the period values are inconsistent with each other, and both disagree with 0.688134(28)~days from \citet{str14}, by 0.45\% and 0.23\% respectively.  The Strader et al. spectroscopic ephemeris is clearly correct because it continues to accurately fit the phases of our light curves 7~yr later, while the periods from Pal et al. produce a phase drift with respect to this standard, by either 0.17 cycles or 0.34 cycles over the 52~day span of the TESS observation alone.

\begin{figure}
\vspace{-0.2in}
  \centerline{
\includegraphics[angle=0.,width=1.05\linewidth]{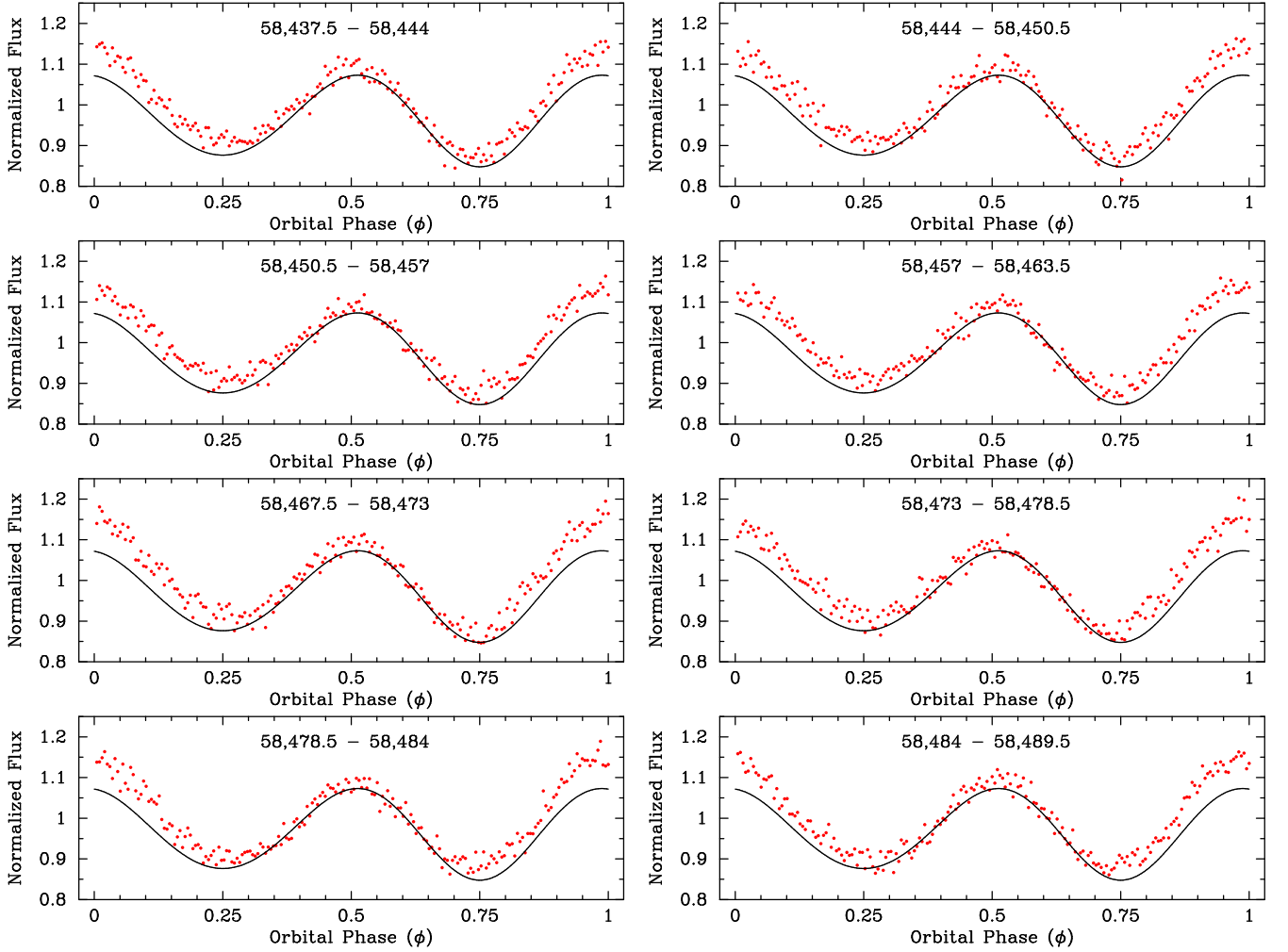}
  }
\vspace{-0.2in}
  \caption{Binned and folded TESS light curves (according to the ephemeris of \citealt{str14}), labeled with their MJD dates.  The ellipsoidal model of Figure~\ref{fig:model} is shown for comparison.}
\label{fig:tess}
\end{figure}

  We analyzed the TESS light curve files in order to check the ephemeris and search for any variability of the type seen in the MDM light curves.  A Lomb-Scargle periodogram gives a period from TESS of 0.68751(28)~days using the fundamental and 0.688109(34)~days using the harmonic.  The latter is fully consistent with Strader et al., while the less precise fundamental is only marginally in agreement.  The small differences may be due to evolution of the waveform during the span of the observation.

In Figure~\ref{fig:tess}, we show binned light curves of the TESS data divided into segments of 6.5 or 5.5~days, folded on the Strader et al. ephemeris.  This shows essentially perfect agreement in phase, as already seen in the MDM observations.  The ellipsoidal model of Figure~\ref{fig:model} is reproduced for comparison.  Some of the same deviations that are seen in the MDM light curves are present here, such as the differing heights of the maxima, and phase offsets of the minima.  However, we emphasize that these effects are not variable enough to allow the Pal et al. periods.  Finally, we also examined unbinned light curves and find no obvious flares.

\begin{figure}[!ht]
\vspace{-7.7in}
\hspace{-0.2in}
\centerline{
\includegraphics[width=1.25\linewidth]{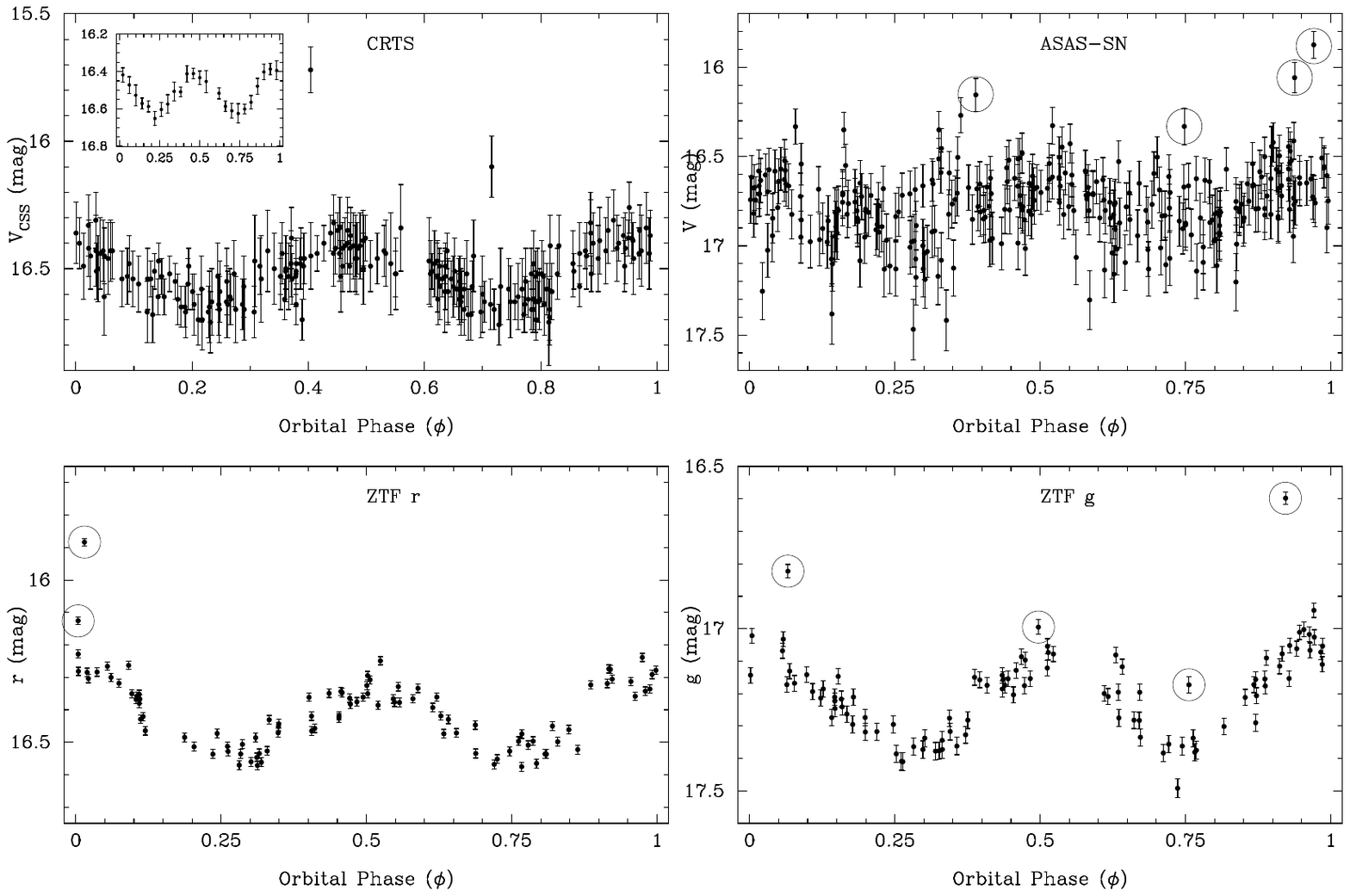}
}
\caption{
Optical survey data on \obj\ folded on the orbital ephemeris of \citet{str14}. 
Top left: CRTS data from the years 2006--2013, excluding points with error bars
$>0.2$~mag.  The two high points with $V_{\rm CSS}=16.10$ and 15.72 are candidate flares.
The inset is a binned folded light curve, excluding the two high points. Top right: ASAS-SN $V$-band data from 2014--2018.  Bottom: ZTF $g$- and $r$-band data from 2018--2021.  Likely flare points are circled. Times of these points are listed in Table~\ref{tab:flares}.} 
\label{fig:surveys}
\end{figure}

\section{Archival Optical Surveys\label{sec:surveys}}

We examined the same data that \citet{str14} used from the CRTS second data release, collected over the years 2006--2013.   Figure~\ref{fig:surveys} shows the data folded on the orbital period, excluding points whose error bars are $>0.2$~mag, leaving 229 points.  Evidently there are two high points, which we list in Table~\ref{tab:flares} as candidate flares.  On a typical night, four unfiltered 30~s exposures of a field are taken in a time span of $\sim30$ minutes.  This was the case with the candidate flares, but in each case only one point stands out, which raises suspicion that they are spurious.  However, the nearest neighboring point is separated by $>15$~minutes, so it is conceivable that a short flare such as the ones detected on 2021 January 10 in Figure~\ref{fig:time-series}f was caught in only one exposure.

We also used the ZTF 10th data release, which provided 101 points in $r$ and 102 in $g$ over the years 2018--2021. Six high points in the folded ZTF light curves in Figure~\ref{fig:surveys} are marked, which come from three different flaring episodes as listed in Table~\ref{tab:flares}.  In two of these episodes, more than one high point was detected within one day; therefore, we consider these multiple detections to be confirmed flares.  In fact, the event of 2020 January 23--24 was detected at the same time that MDM was observing it (Figure~\ref{fig:time-series}b).  There may be additional low-level flaring episodes that contribute to the excess scatter of the ZTF points over a smooth curve, but we have not attempted to identify all such candidates in these data.  It also appears that the minimum at $\phi=0.25$ is late, at least in most of the ZTF data.

ASAS-SN data comprise 314 detections in $V$ over the years 2014--2018, filling the gap between CRTS and ZTF.  We list four points as probable flares that are $\ge0.5$~mag brighter than the quiescent level.  As in the other surveys, there may be lower-level flares in these data as well.  Two of the four flares in ASAS-SN occurred within 17 days of each other in 2015 September.  Eight intervening measurements between these two were of normal brightness, so this not necessarily indicative of a duty cycle for a flaring episode.  Such a coincidence may occur over four seasons of monitoring.

\begin{deluxetable}{ccccc}[h]
\label{tab:flares}
\tablecolumns{5} 
\tablewidth{0pt} 
\tablecaption{Flares in CRTS, ASAS-SN, and ZTF Data} 
\tablehead{
\colhead{Band} & \colhead{Day (BJD)} & \colhead{Date/Time (UTC)} & 
\colhead{Magnitude} & \colhead{Phase ($\phi$)}
}
\startdata
 $V_{\rm CSS}$  &   2454320.32881  &  2007-08-07 19:53:23  &  15.72(9)   &  0.404 \\
\hline
 $V_{\rm CSS}$  &   2454899.95191  &  2009-03-09 10:49:46  &  16.10(12)  &  0.706 \\
\hline
 $V_{\rm ASAS-SN}$  &   2457276.91974  &  2015-09-11 10:03:12  & 16.058(86)  &  0.938  \\
\hline
 $V_{\rm ASAS-SN}$  &   2457293.74568  &  2015-09-28 05:51:00  & 16.154(92)  &  0.389  \\
\hline
 $V_{\rm ASAS-SN}$  &   2457409.75224  &  2016-01-22 05:58:10  &  15.874(75) &  0.971 \\
\hline
 $V_{\rm ASAS-SN}$  &   2458148.65494  &  2018-01-30 03:38:36  &  16.33(10) &  0.748 \\
\hline
 $g_{\rm ZTF}$  &   2458767.98108  &  2019-10-11 11:28:34  &  17.173(25)  &  0.756 \\
\hline
 $g_{\rm ZTF}$  &   2458871.71100  &  2020-01-23 04:58:35  &  16.995(23)  &  0.497 \\
 $g_{\rm ZTF}$  &   2458872.69188  &  2020-01-24 04:31:08  &  16.598(20)  &  0.922 \\
 $r_{\rm ZTF}$  &   2458872.75550  &  2020-01-24 06:02:45  &  15.883(14)  &  0.015 \\
\hline
 $r_{\rm ZTF}$  &   2459216.81502  &  2021-01-02 07:27:16  &  16.126(14)  &  0.004 \\
 $g_{\rm ZTF}$  &   2459216.85765  &  2021-01-02 08:28:40  &  16.823(21)  &  0.066 \\
\enddata
\end{deluxetable}

\section{Swift X-ray Observations\label{sec:xrt}}

\begin{deluxetable}{llcrc}
\label{tab:xraylog}
\tablecolumns{5} 
\tablewidth{0pt} 
\tablecaption{Log of Swift XRT Observations}
\tablehead{
\colhead{ObsID} & \colhead{Date} & \colhead{Time} & \colhead{Exposure} &
\colhead{Mean Count Rate} \\
& \colhead{(UT)} & \colhead{(UTC)} & \colhead{(s)} & \colhead{(s$^{-1}$)}
}
\startdata
00031535001\tablenotemark{a} & 2009 Nov 12     & 01:01--12:23   &  4791 &  $(2.4\pm0.9)\times10^{-3}$ \\
00032938001\tablenotemark{a} & 2013 Sep 17    & 02:17--23:33   & 14440 &  $(5.4\pm0.7)\times10^{-3}$ \\
00092233001\tablenotemark{a}	& 2016 Sep 15    & 21:18--21:36   &  1078 &  $(2.8\pm1.9)\times10^{-3}$ \\
00092233002\tablenotemark{a}	& 2016 Oct 20     & 19:50--22:10   &  1080 &  $(4.5\pm2.3)\times10^{-3}$ \\
00092233003\tablenotemark{a}	& 2016 Nov 24     & 01:42--01:58   &   978 &  $(5.7\pm2.8)\times10^{-3}$ \\
00092233004\tablenotemark{a}	& 2016 Dec 29     & 14:29--16:12   &  1009 &  $(6.4\pm22)\times10^{-4}$ \\
00092233005\tablenotemark{a}	& 2017 Feb 2      & 22:49--23:05   &   995 &  $(5.4\pm4.7)\times10^{-3}$ \\
00092233006	& 2017 Mar 9      & 04:09--04:26   &  1046 &  $(1.9\pm0.5)\times10^{-2}$ \\
00013184001 & 2020 Jan 26     & 00:52--10:56   &  4837 &  $(1.4\pm0.2)\times10^{-2}$ \\
00013184002\tablenotemark{b} & 2020 Jan 27     & 11:44--14:01   &  2653 &  $(4.1\pm0.4)\times10^{-2}$ \\
00013184003 & 2020 Jan 28     & 03:55--21:28   &  4523 &  $(1.3\pm0.2)\times10^{-2}$ \\
00013184004 & 2020 Jan 29     & 14:59--23:11   &  3790 &  $(1.2\pm0.3)\times10^{-2}$ \\
00013184005\tablenotemark{a} & 2020 Feb 24     & 01:10--14:03   &  4697 &  $(4.5\pm1.3)\times10^{-3}$ \\
00013184006\tablenotemark{a} & 2020 Feb 25     & 01:10--22:04   &  2027 &  $(2.5\pm13)\times10^{-4}$ \\
00013184007\tablenotemark{a} & 2020 Feb 26     & 18:37--23:34   &  2408 &  $(5.2\pm2.0)\times10^{-3}$ \\
00013184008\tablenotemark{a} & 2020 Feb 27     & 07:14--23:38   &  3399 &  $(5.8\pm1.5)\times10^{-3}$ \\
00013184009 & 2020 Mar 3      & 00:24--00:49   &  1509 &  $(1.3\pm0.3)\times10^{-2}$ \\
00013184010\tablenotemark{a} & 2020 Mar 25     & 09:21--17:53   &  4618 &  $(3.7\pm2.2)\times10^{-3}$ \\
00013184011\tablenotemark{a} & 2020 Mar 26     & 06:11--09:42   &  3523 &  $(6.2\pm1.7)\times10^{-3}$ \\
00013184012 & 2020 Mar 27--28 & 01:08--06:13   &  3737 &  $(1.3\pm0.2)\times10^{-2}$ \\
00013184013 & 2020 Apr 1      & 04:05--11:57   &  1549 &  $(2.0\pm0.4)\times10^{-2}$ \\
00013184014\tablenotemark{c} & 2020 Apr 24--25 & 20:41--05:01   &  6275 &  $(1.5\pm0.2)\times10^{-2}$ \\
00013184015\tablenotemark{c} & 2020 Apr 26--27 & 15:42--17:26   & 10948 &  $(1.3\pm0.1)\times10^{-2}$ \\
00013184016\tablenotemark{c} & 2020 Apr 29     & 02:39--09:13   &  6792 &  $(1.2\pm0.2)\times10^{-2}$ \\
00095619001\tablenotemark{b} & 2020 May 10     & 03:34--06:47   &  2084 &  $(3.1\pm0.4)\times10^{-2}$ \\
00095619002\tablenotemark{b} & 2020 Aug 2      & 01:11--01:25   &   842 &  $(3.1\pm0.8)\times10^{-2}$ \\
00095619003\tablenotemark{a} & 2020 Oct 25     & 13:39--13:57   &  1078 &  $(5.5\pm2.8)\times10^{-3}$ \\
00095619004 & 2021 Jan 17     & 08:37--08:53   &   953 &  $(1.1\pm0.4)\times10^{-2}$ \\
00095807001\tablenotemark{a} & 2021 Nov 28     & 12:13--12:28   &   910 &  $(7.2\pm3.1)\times10^{-3}$ \\
\enddata
\tablenotetext{a}{Quiescent state in  Table~\ref{tab:xspec}: observations with mean count rate $<1.0\times10^{-2}$~s$^{-1}$.}
\tablenotetext{b}{Flare state in  Table~\ref{tab:xspec}: observations with mean count rate $>3.0\times10^{-2}$~s$^{-1}$.}
\tablenotetext{c}{Intermediate state in  Table~\ref{tab:xspec}: selected observations with long exposure times and mean count rate $(1.2-1.5)\times10^{-2}$~s$^{-1}$.}
\end{deluxetable}

\begin{figure}
\includegraphics[scale=0.7]{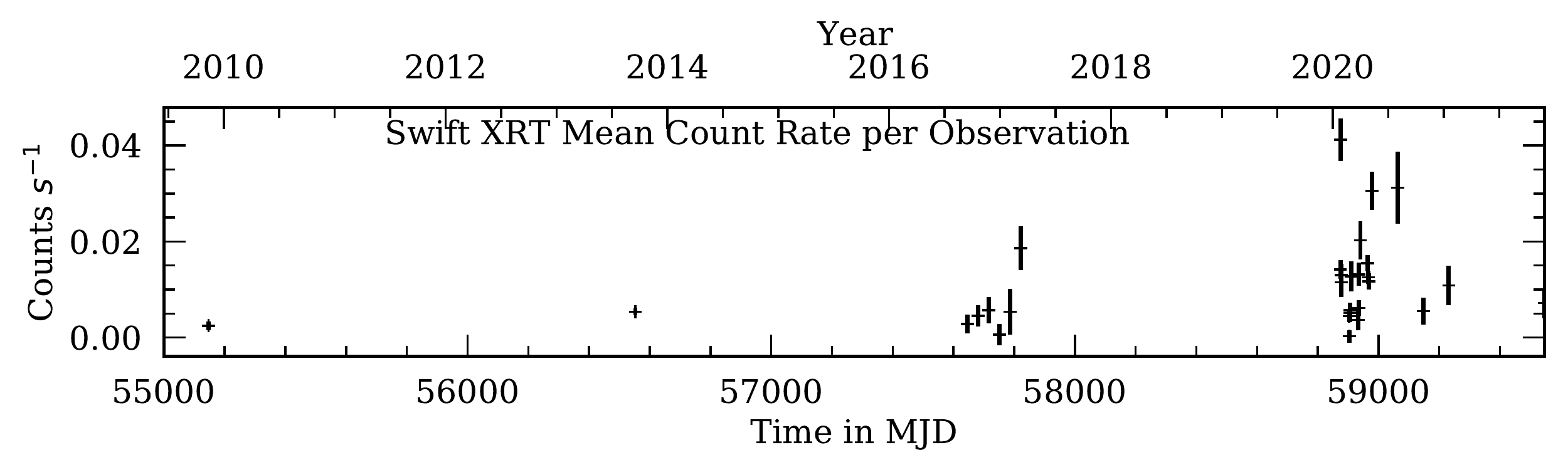}
\caption
{Mean count rate (0.3--10 keV) per observation in the Swift XRT, from Table~\ref{tab:xraylog}.}
\label{fig:XRT_allobs}
\end{figure}

Table~\ref{tab:xraylog} is a log of all observations of \obj\ by the Swift XRT through the end of 2021, while Figure~\ref{fig:XRT_allobs} plots their mean count rates, which are highly variable.
\citet{xin14} analyzed the first two observations (in 2009 and 2013), and found a mean count rate of $3.7\times10^{-3}$~s$^{-1}$, corresponding to an average flux of $1.3\times10^{-13}$ erg~cm$^{-2}$~s$^{-1}$ in the 0.3--7~keV band.
Their fitted photon spectral index was $\Gamma=1.5\pm0.2$.
Using the Gaia EDR3 parallax of $0.446\pm0.041$ indicating a nominal distance of 2.24~kpc, the X-ray luminosity was then $7.8\times10^{31}$ erg~s$^{-1}$.  In comparison with later observations, this evidently represents the ``quiescent'' luminosity of \obj, which is typical of redbacks \citep{lin14,str19}.

In 2017 March a significantly higher count rate of $(1.9\pm0.5)\times10^{-2}$ was detected, but we did not observe \obj\ again until 2020, when we triggered Swift observations immediately following our discovery of optical flaring.  Fast flaring was then detected in several of these triggered observations, which we analyzed as described next.

\subsection{X-ray Light Curves}
\label{sect:lightcurves}

\begin{figure}
\centerline{
\includegraphics[scale=0.7]{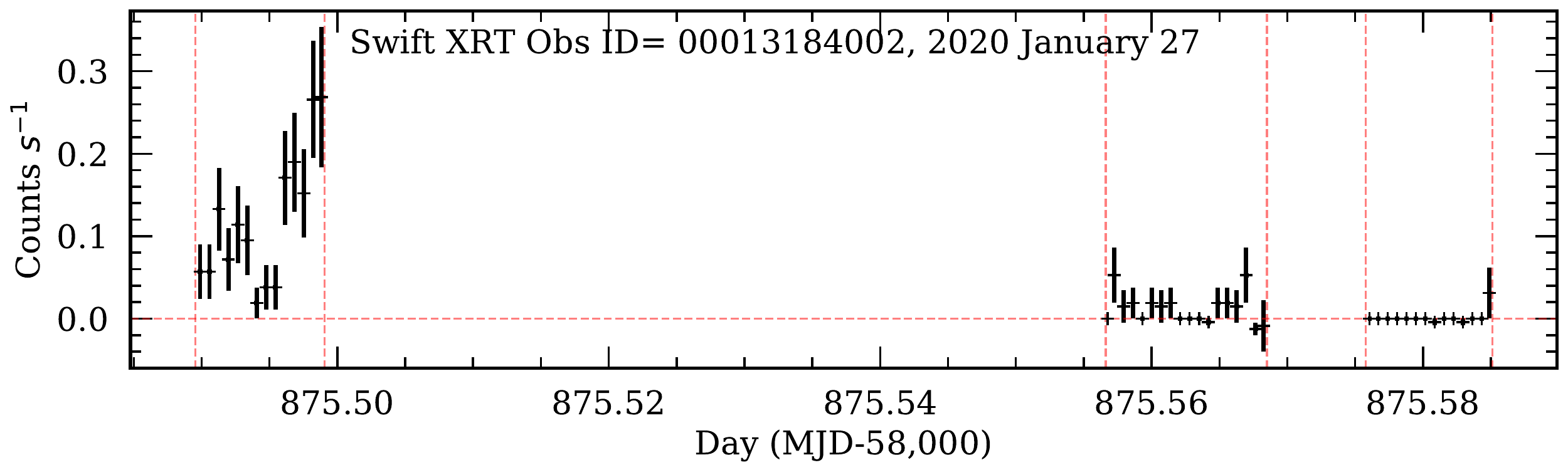}
}
\vspace{-0.2in}
\centerline{
\includegraphics[scale=0.665]{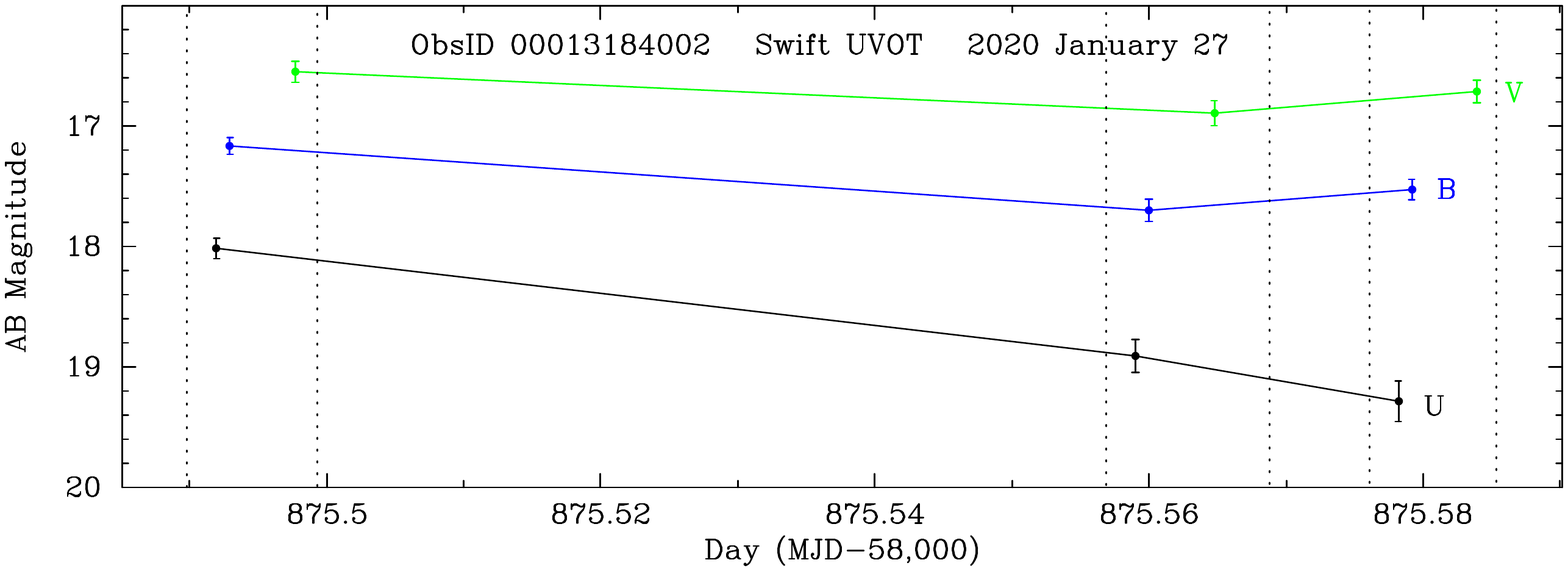}
}
\vspace{-3.15in}
\centerline{
\includegraphics[scale=0.665]{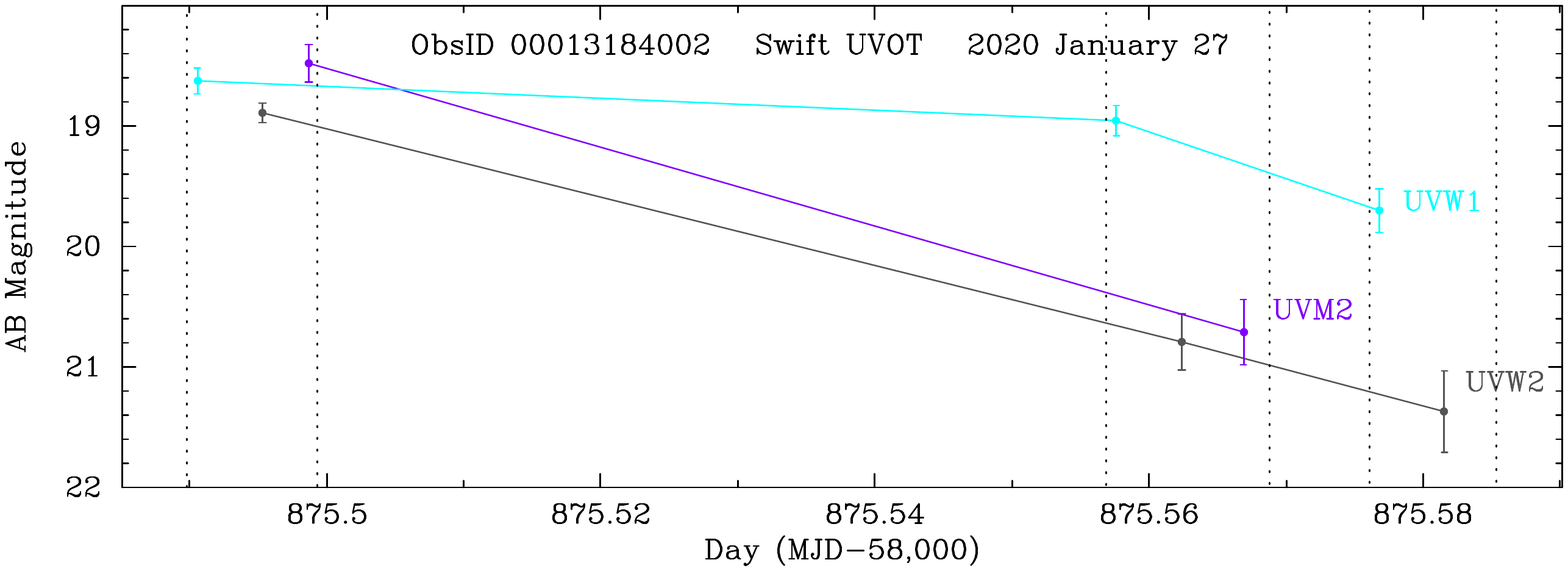}
}
\vspace{-3.in}
\caption
{
Bright flare seen by Swift on 2020 January 27. Top panel: XRT light curve in 60~s bins.  Dotted vertical lines bound the GTIs. Middle panel: $U,B,V$ magnitudes.  Bottom panel: $\mathit{UVW1}, \mathit{UVM2}, \mathit{UVW2}$ magnitudes.
}
\label{fig:swift_1}
\end{figure}

\begin{figure}
\centerline
{
\includegraphics[scale=0.7]{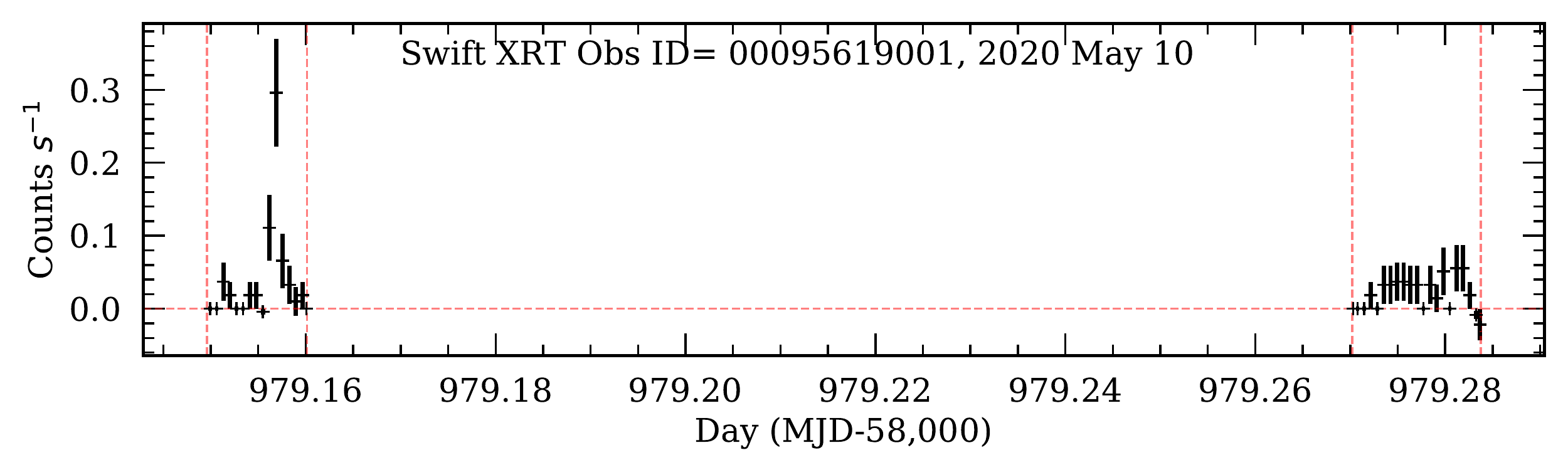}
}
\vspace{-0.3in}
\centerline{
\includegraphics[scale=0.665]{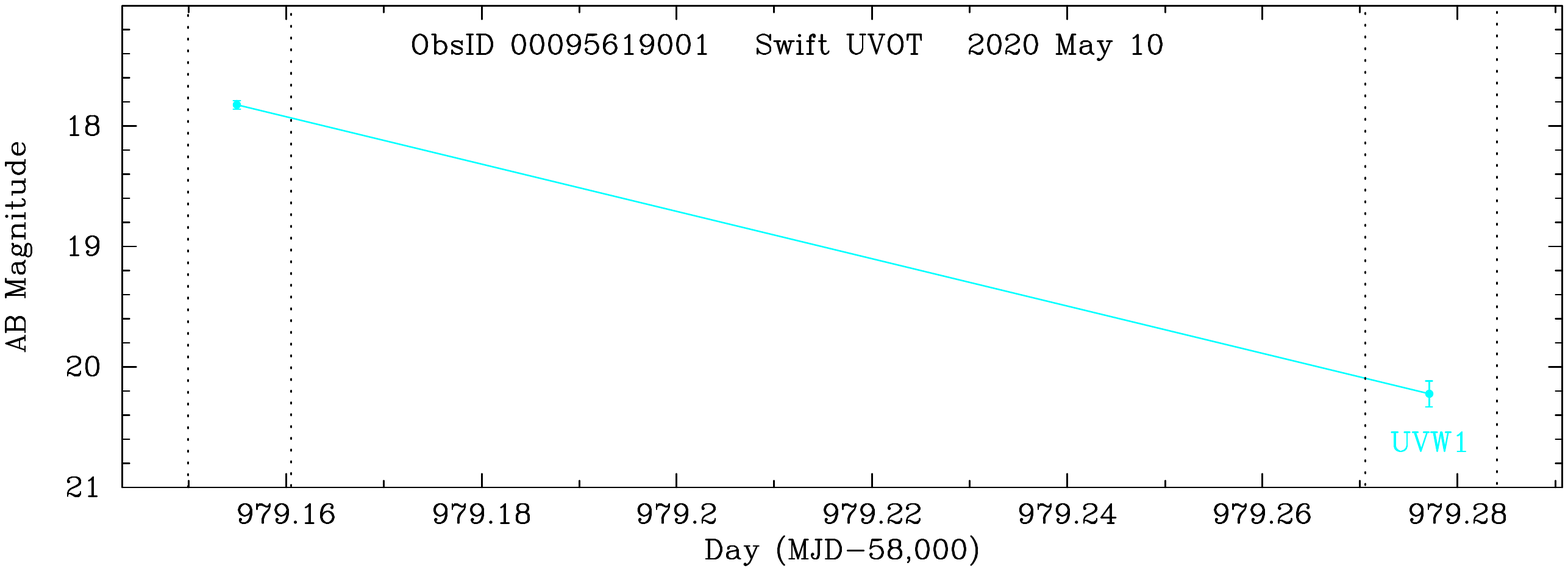}
}
\vspace{-2.96in}
\centerline{
\includegraphics[scale=0.713]{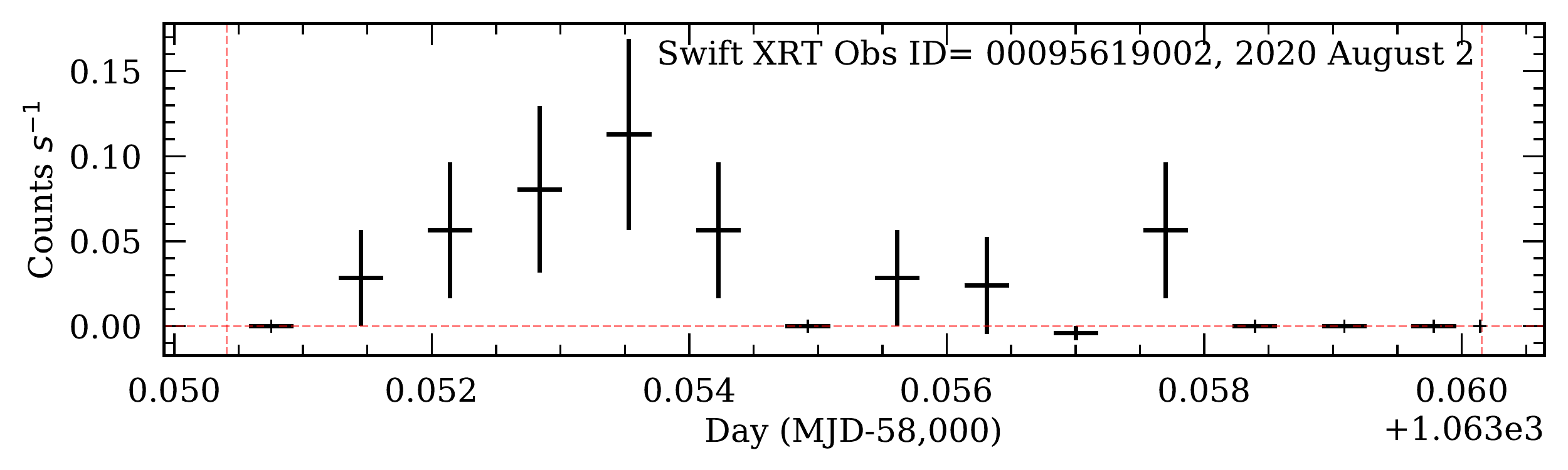}
}
\vspace{-0.1in}
\caption
{
Bright flares seen by Swift on 2020 May 10 and August 2. Top panel: XRT light curve on May 10 in 60~s bins. Middle panel: $\mathit{UVW1}$ magnitudes on May 10.  Bottom panel: XRT light curve on August 2.  The single accompanying $\mathit{UVW1}$ magnitude of 18.30 (see Table~\ref{tab:uvflares}) is not shown.
}
\label{fig:swift_2}
\end{figure}

Using FTOOLS version 6.28, we processed each XRT observation of \obj\ taken in photon counting mode, and produced background-subtracted light curves in the 0.3--10 keV band.  We used an aperture radius of of 25 pixels (59$\arcsec$) for the source and 50 pixels (118$\arcsec$) for the background region.  Within the good time intervals (GTIs), we found the correction factors for bad columns in each 10~s time interval, and applied them to the 60~s time bins of the light curve. 
Using the 60~s bins, we subtracted the background rate from the corrected source rate.  The mean net count rate for each observation is what is listed in Table \ref{tab:xraylog}. 

Beyond the highly variable mean count rates,
several of the observations contain strong flares on time scales of minutes, similar to the optical flares, with the count rate in a 60~s bin reaching as high as 0.3~s$^{-1}$, a factor of 100 above the minimum flux level. Three of these flares are shown in Figures~\ref{fig:swift_1} and \ref{fig:swift_2}. Accompanying the X-ray flares are brightening in the UVOT filters, also shown in the Figures. They will be discussed in Section~\ref{sec:uvot}.

The highest instantaneous XRT count rate of 0.3~s$^{-1}$ occurred in ObsID 00013184002 on 2020 January 27, and ObsID 00095619001 on 2020 May 10, corresponding to a 0.3--10~keV flux of $1.3\times10^{-11}$ erg~cm$^{-2}$~s$^{-1}$, and $L_x \approx 8\times10^{33}$ erg~s$^{-1}$ (see Section~\ref{sec:xspec}).  Note from Figure~\ref{fig:swift_1} that the flare on January 27 may have peaked even higher, as the X-ray flux was possibly still rising at the end of the first GTI.  These are the most luminous X-ray flares that we know of from a non-accreting redback. In comparison, the strong X-ray flares seen in \msptwo\ \citep{cho18} and \fgl\ \citep{hal17b} had peak luminosities of $\approx1\times10^{33}$ erg~s$^{-1}$.  

\newpage
\subsection{X-ray Spectra\label{sec:xspec}}

In order to obtain sufficient statistics for spectral fitting at the various flux levels of \obj, we first grouped observations having similar mean count rates into what we call flaring, intermediate, and quiescent states. The observations selected for these groups are indicated in Table \ref{tab:xraylog}. For each group, we combined the exposure maps using {\fontfamily{qcr}\selectfont ximage}. For each state we combined the same clean Level 2 event files that were used in Section \ref{sect:lightcurves}, and extracted a spectrum from the final image using the extraction regions used in our light curves. We generated the ancilliary response file using the tool {\fontfamily{qcr}\selectfont xrtmkarf}, which corrects for hot columns and bad pixels, and applied the point-spread-function correction.  We used {\fontfamily{qcr}\selectfont grppha} to bin the counts in energy, ignoring energies $<0.3$~keV. We used {\fontfamily{qcr}\selectfont group min 20} to ensure that there are at least 20 counts per bin in order to employ $\chi^{2}$ statistics.  We grouped any remaining counts into a final bin, which therefore has an upper energy limit defined by the highest energy photon; energy channels above that are ignored. Using XSPEC, we fit absorbed power-law models and integrated their flux over a 0.3--8 keV range.  Because the column density is small and weakly constrained by the fits, we also computed models in which $N_{\rm H}$ is fixed at $1.48\times10^{20}$~cm$^{-2}$, the total Galactic value along the line of sight as given by \citet{HI4PI}.  At its high Galactic latitude of $-29^{\circ}\!.8$, \obj\ is distant enough for the the total Galactic $N_{\rm H}$ to apply.

\begin{deluxetable}{lclccc}[ht]
\label{tab:xspec}
\tablecolumns{6}
\tablewidth{0pt}
\tablecaption{Swift XRT Spectral Fits for Selected States}
\tablehead{
    \colhead{State\tablenotemark{a}} & \colhead{Exposure} & \colhead{$N_{\rm H}$} & \colhead{$\Gamma$} & \colhead{$F_X$\tablenotemark{b}} & \colhead{$\chi^2$/dof} \\
    & \colhead{(s)} & \colhead{($10^{20}$~cm$^{-2}$)} & & \colhead{(0.3--8 keV)} &
    }
    \startdata
    Flare & 5579 & $6.85^{+12.05}_{-6.87}$ & $1.87^{+0.38}_{-0.31}$ & $14.59^{+0.32}_{-0.24}$ & 2.54/7 \\
    &  & 1.48\tablenotemark{c} & $1.70^{+0.17}_{-0.16}$ & $13.70^{+0.20}_{-0.19}$ & 3.62/8 \\
    \hline
    Intermediate & 18015 & $2.24^{+7.21}_{-2.22}$ & $1.39^{+0.23}_{-0.18}$ & $5.95^{+0.71}_{-0.70}$ & 21.47/14 \\
    &  & 1.48\tablenotemark{c} & $1.37^{+0.14}_{-0.14}$ & $5.93^{+0.71}_{-0.69}$ & 21.52/15 \\
    \hline
    Quiescent & 47031 & $3.94^{+12.56}_{-3.94}$ & $1.51^{+0.37}_{-0.28}$ & $1.97^{+0.32}_{-0.31}$ & 11.25/10 \\
    & & 1.48\tablenotemark{c} & $1.44^{+0.20}_{-0.20}$ & $1.95^{+0.33}_{-0.30}$ & 11.45/11 \\
    \enddata
    \tablecomments{Errors are 90\% confidence for one interesting parameter.}
    \tablenotetext{a}{See Table~\ref{tab:xraylog} for included observations.}
    \tablenotetext{b}{Unabsorbed 0.3--8~keV flux in units of $10^{-13}$ erg cm$^{-2}$ s$^{-1}$.}
    \tablenotetext{c}{$N_{\rm H}$ held fixed at the Galactic value from \citet{HI4PI}.}
\end{deluxetable}

The results in Table~\ref{tab:xspec} show that the quiescent state of \obj, represented by the 15 faintest observations in Table~\ref{tab:xraylog}, has a mean 0.3--8~keV luminosity of $1.2\times10^{32}$ erg~s$^{-1}$.  This is similar to other redbacks \citep{str19}.  The intermediate state has a mean luminosity of $4.2\times10^{32}$ erg~s$^{-1}$.  The flaring states are on average another factor of 2--3 higher, but the three observations in this group are highly variable in themselves. To characterize the peak instantaneous luminosity of a flaring state, we apply a counts-to-erg conversion factor of $4.0\times10^{-11}$ erg~count$^{-1}$, derived from the average flare spectrum, to the light curves in Figures~\ref{fig:swift_1} and \ref{fig:swift_2}.  The maximum observed count rate of $\approx0.3$~s$^{-1}$ corresponds to a luminosity of $8\times10^{33}$ erg~s$^{-1}$.  The total range of luminosity displayed by quiescent and flaring observations is thus a factor of $\approx 100$, as was seen in the light curves.

Another result is that the photon index of $\Gamma = 1.6-1.9$ in the flaring state is steeper than $\Gamma=1.4-1.5$ in the quiescent state.  This is the opposite of what was found in the case of the flaring redback \msptwo, in which the quiescent state was similar to that of \obj,  while the flaring spectrum was even flatter, with $\Gamma\approx 1.2$ \citep{cho18}.  However, the intermediate state of \obj\ does have a fairly flat spectrum, with $\Gamma = 1.3-1.4$.

\section{Swift UVOT Observations\label{sec:uvot}}
Along with the XRT, \obj\ was observed with the UVOT through all of its filters ($V, B, U, \mathit{UVW1}, \mathit{UVM2}$, and $\mathit{UVW2}$). The photometry was extracted using the \verb|uvotsource| command in FTOOLS. The net flux was obtained from an aperture of $5\arcsec$ radius for the source and $20\arcsec$ for the background region.  Magnitudes are in the AB system \citep{bre11}.  Table~\ref{tab:uvotlog} is a log of the observations, including the number of exposures in each filter.  Many observations had multiple exposures per filter, which were examined individually to look for variability.  Some exposures yielded only upper limits; the number of these is indicated in parentheses. We chose the threshold for reporting a detection as $3\sigma$.

\begin{figure}
  \vspace{-4.in}
  \centerline{
  \hspace{-0.25in}
\includegraphics[width=1.2\linewidth]{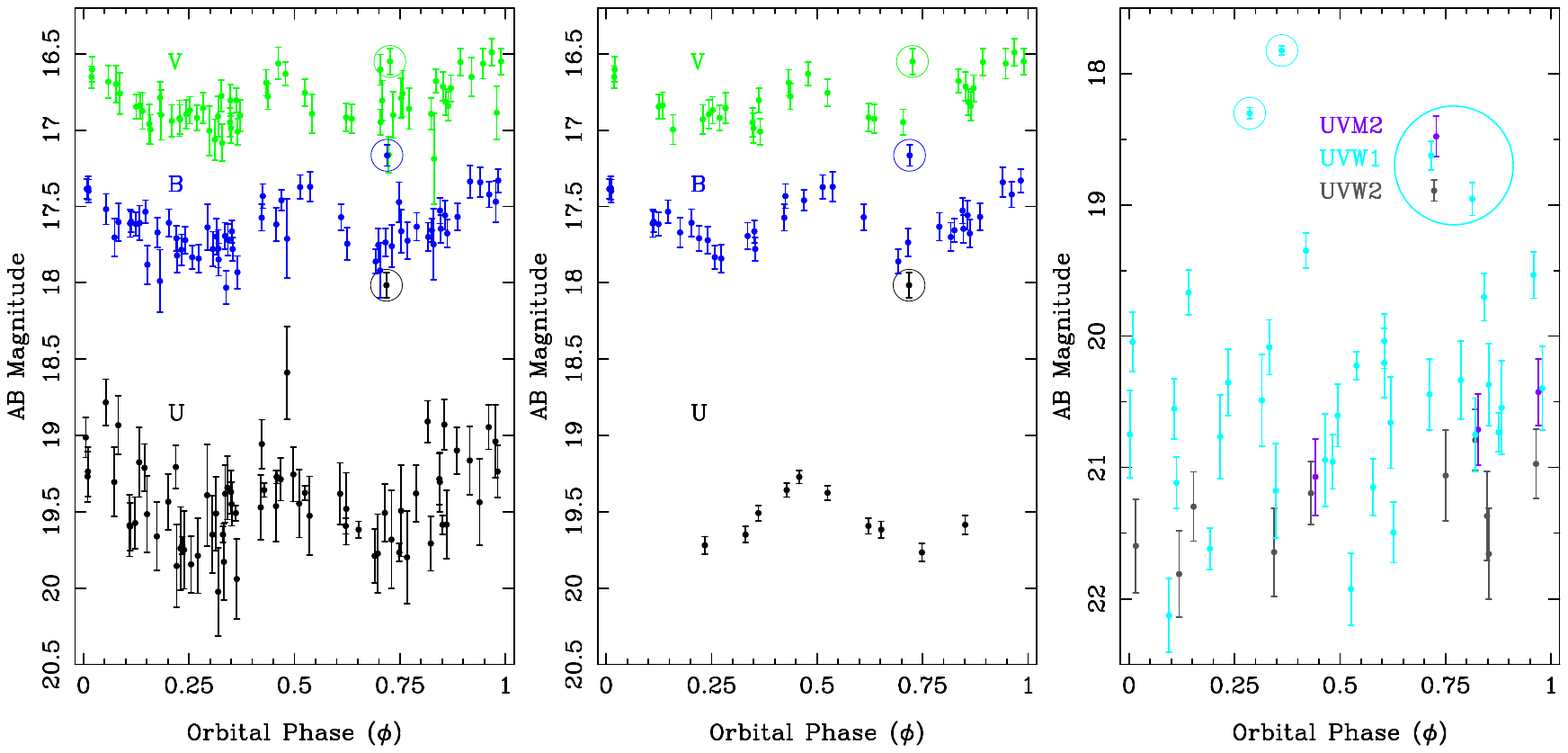}
  }
\vspace{-4.5in}
\caption{Swift UVOT magnitudes from all observations listed in Table~\ref{tab:uvotlog}, folded according to orbital phase.  Only points with $\ge3\sigma$ detection significance are shown.  Left panel: All $U,B,V$ detections.  Middle panel: The subset of $U,B,V$ points with uncertainties $\le0.1$~mag.  Right panel: UV filter magnitudes.  Circled points in each panel are the peaks of flares associated with the three brightest X-ray flares shown in Figures~\ref{fig:swift_1} and \ref{fig:swift_2}. See Table~\ref{tab:uvflares} for details of these brightest UVOT flare data. 
}
\label{fig:uvot}
\end{figure}

All of the UVOT detections are plotted in Figure~\ref{fig:uvot}, folded according to orbital phase.  This shows the normally dominant effect of ellipsoidal modulation on the flux in $U, B$, and $V$ as expected.  The middle panel of Figure~\ref{fig:uvot} shows only the $U,B,V$ points with uncertainties $\le0.1$~mag in order clarify the effect.  The amplitude of modulation increases toward shorter wavelength, as would be expected for the effects of limb darkening and gravity darkening.  In contrast, detections in the UV filters are sparse, and have no obvious dependence on orbital phase.  Most of the UV detections may be due to flares, not quiescent photospheric emission.

\begin{deluxetable}{llcccccc}
\label{tab:uvotlog}
\tablecolumns{5} 
\tablewidth{0pt} 
\tablecaption{Log of Swift UVOT Photometry}
\tablehead{
\colhead{ObsID} & \colhead{Date (UT)} & \multicolumn6{c}{Number of Images\tablenotemark{a}} \\
&  & \colhead{$V$} & \colhead{$B$} & \colhead{$U$} & \colhead{$\mathit{UVW1}$} & \colhead{$\mathit{UVM2}$} & \colhead{$\mathit{UVW2}$}
}
\startdata
00031535001 & 2009 Nov 12 & -- & -- & -- & 4 & -- & --  \\
00032938001 & 2013 Sep 17 & -- & -- & 9 & -- & -- & -- \\
00092233001 & 2016 Sep 15 & -- & -- & -- & -- & 1(1) & -- \\
00092233002 & 2016 Oct 20 & -- & -- & -- & -- & -- & 2(2) \\
00092233003 & 2016 Nov 24 & -- &  -- & 1 & -- & -- & --  \\
00092233004 & 2016 Dec 29 & -- & -- & -- & 2 & -- & --  \\
00092233005 & 2017 Feb 2 & -- & -- & -- & -- & 1(1) & --  \\
00092233006 & 2017 Mar 9 & -- & -- & -- & -- & -- & 1 \\
00013184001 & 2020 Jan 26 & 4(1) & 6 & 6 & 7(3) & 3(3) & 6(6) \\
00013184002 & 2020 Jan 27 & 3 & 3 & 3 & 3 & 3(1) & 3 \\
00013184003 & 2020 Jan 28 & 7 & 7 & 7 & 7(7) & 7(7) & 7(7) \\
00013184004 & 2020 Jan 29 & 3 & 3 & 3 & 3(1) & 3(3) & 3(2) \\
00013184005 & 2020 Feb 24 & 6 & 6 & 6 & 6(5) & 6(6) & 6(6) \\
00013184006 & 2020 Feb 25 & 3 & 4 & 4 & 4(4) & 3(3) & 4(4) \\
00013184007 & 2020 Feb 26 & 3 & 3 & 3(1) & 4(3) & 4(4) & 4(4) \\
00013184008 & 2020 Feb 27 & 2 & 2 & 3 & 3(1) & 2(2) & 2(2) \\
00013184009 & 2020 Mar 3 & 1 & 1 & 1 & 1(1) & 1(1) & 1(1) \\
00013184010 & 2020 Mar 25 & 3 & 3 & 3 & 3(2) & 3(3) & 3(3) \\
00013184011 & 2020 Mar 26 & 3 & 3 & 3 & 3(2) & 3(3) & 3(3) \\
00013184012 & 2020 Mar 27--28 & 4 & 4 & 4 & 4(2) & 3(2) & 4(2) \\
00013184013 & 2020 Apr 1 & 2 & 2 & 2(1) & 2(1) & 2(2) & 2(2) \\
00013184014 & 2020 Apr 24--25 & 4 & 4 & 4 & 4 & 4(4) & 4(2) \\
00013184015 & 2020 Apr 26--27 & 8 & 8 & 8(1) & 8(7) & 8(7) & 8(5) \\
00013184016 & 2020 Apr 29 & 5 & 5 & 5(2) & 5(3) & 5(5) & 5(5) \\
00095619001 & 2020 May 10 & -- & -- & -- & 2 & -- & -- \\
00095619002 & 2020 Aug 2 & -- & -- & -- & 1 & -- & -- \\
00095619003 & 2020 Oct 25 & -- & -- & -- & 1 & -- & -- \\
00095619004 & 2021 Jan 17 & -- & -- & -- & 1 & -- & -- \\
00095807001 & 2021 Nov 28 & -- & -- & -- & -- & 1(1) & -- \\
\enddata
\tablenotetext{a}{Number of images in each filter. The number of these yielding non-detections (upper limits) is in parentheses.}
\end{deluxetable}

It is clear that the bright X-ray flares are accompanied by brightening in optical and UV, and the amplitude of the flare increases to shorter wavelengths.  Table~\ref{tab:uvflares} lists the UVOT magnitudes that accompanied the three brightest XRT flares, as shown in Figures~\ref{fig:swift_1} and \ref{fig:swift_2}.  The peak magnitudes of these flares are also circled in Figure~\ref{fig:uvot}.  In the $V$ and $B$ bands the stellar photosphere makes a significant contribution to the flux even at the peak of the flare, while in the $U$ band the flux during the brightest flare increased by $\approx 1.7$~mag.  In the UV bands the increase is $2-4$~mag.  The brightest magnitude of 17.84 in the $\mathit{UVW1\/}$ filter on 2020 May 10 corresponds to an extinction-corrected luminosity of $\nu L_{\nu}=1.6\times10^{33}$~erg~s$^{-1}$.  See Section \ref{sec:disc_c} for extinction corrections and the spectral energy distribution of the flare of 2020 January 27.

\begin{deluxetable}{llcclc} 
\label{tab:uvflares}
\tablecolumns{6} 
\tablewidth{0pt} 
\tablecaption{Bright Flare Data in the Swift UVOT}
\tablehead{
\colhead{ObsID} & \colhead{Date} & \colhead{Day (MJD)\tablenotemark{a}} & \colhead{Filter} & 
\colhead{Magnitude} & \colhead{Phase ($\phi$)}
}
\startdata
00013184002  & 2020 Jan 27 &  58875.4953  &  $\mathit{UVW2}$  &  18.891(80)  &  0.723 \\
            & &  58875.5624  &  $\mathit{UVW2}$  &  20.79(23)   &  0.820 \\
            & &  58875.5815  &  $\mathit{UVW2}$  &  21.37(34)   &  0.848 \\
            & &  58875.4987  &  $\mathit{UVM2}$  &  18.48(16)   &  0.728 \\
            & &  58875.5669  &  $\mathit{UVM2}$  &  20.71(27)   &  0.827 \\
            & &  58875.4905  &  $\mathit{UVW1}$  &  18.63(11)   &  0.716 \\
            & &  58875.5576  &  $\mathit{UVW1}$  &  18.96(13)   &  0.813 \\
            & &  58875.5768  &  $\mathit{UVW1}$  &  19.70(18)   &  0.841 \\
            & &  58875.4919  &  $U$   &  18.016(84)  &  0.718 \\
            & &  58875.5590  &  $U$   &  18.91(14)   &  0.816 \\
            & &  58875.5782  &  $U$   &  19.29(17)   &  0.843 \\
            & &  58875.4929  &  $B$   &  17.165(70)  &  0.719 \\
            & &  58875.5600  &  $B$   &  17.700(93)  &  0.817 \\
            & &  58875.5792  &  $B$   &  17.527(84)  &  0.845 \\
            & &  58875.4977  &  $V$   &  16.550(86)  &  0.726 \\
            & &  58875.5648  &  $V$   &  16.89(10)   &  0.824 \\
            & &  58875.5839  &  $V$   &  16.714(94)  &  0.852 \\
\hline
00095619001  &  2020 May 10 & 58979.1549  &  $\mathit{UVW1}$  &  17.842(36)  &  0.362 \\
            & &  58979.2779  &  $\mathit{UVW1}$  &  20.22(11)  &  0.539 \\
\hline
00095619002  &  2020 Aug 2 & 59063.0553  &  $\mathit{UVW1}$  &  18.302(44)  &  0.286 \\
\enddata
\tablenotetext{a}{Time of mid-exposure.}
\end{deluxetable}

\section{Discussion\label{sec:disc}}

\subsection{Nature of the Companion Star\label{sec:disc_a}}

\citet{str14} classified the companion as late G or early K based on its optical spectrum.  We use the multicolor photometry from ZTF and Swift, correcting for the modest total Galactic extinction of $A_V = 0.075$ in this high-latitude direction \citep{sch11},
to derive intrinsic apparent magnitudes in the Vega system. Averaged over the ellipsoidal light curve, mean intrinsic values are $V=16.64, U-B=0.70, B-V=0.92$.  The colors correspond most closely to a K2 main sequence star, but a K2 star has absolute magnitude $\approx6.5$ which at a distance of 2.24~kpc would be seen as $V=18.25$. Therefore, the star is larger that a K2 star by a factor of 2.1 in radius.  The orbital parameters derived by \citep{str19} imply a semi-major axis of $3\times10^{11}$~cm and a Roche lobe radius of $1\times10^{11}$~cm, which is 1.8 times the radius of a K2 star, in close agreement with the photometry.  It is common for redback companions to be less dense than a main sequence star of the same mass.

While the optical modulation is mostly ellipsoidal, considerable variation of its shape makes it difficult to use it to diagnose fundamental properties of a binary, such as the Roche lobe filling factor and orbital inclination. To illustrate this, we overlaid an approximate ellipsoidal model on the 2020 November data (Figure~\ref{fig:model}), which is devoid of flares.  The equations of \citet{mor93} were used, together with the correction factors derived by \cite{gom21} that are needed because \cite{mor93} is only valid for Roche-lobe filling factors $f<\!<1$.  The illustrated model is not a fit to the data; rather, the spectroscopically determined orbital parameters from \citet{str14} are used together with limb-darkening and gravity-darkening coefficients appropriate for the spectral class \citep{cla11}.

The basic symmetries of the model that are violated by the data are the unequal maxima at the quadrature points $\phi=0$ (higher) and $\phi=0.5$ (lower), and the separation in phase of the minima, which are not exactly 0.5 apart.  The minimum at inferior conjunction of the secondary, $\phi=0.25$, lags by about 0.05, appearing instead at $\phi\approx0.30$ (if the orbital ephemeris is still valid).  Both of these effects point to an axial asymmetry in which one quadrant of the star away from the L1 point is either dimmer or brighter due to a large starspot, or asymmetric heating, respectively. For example, a dark spot on the trailing, outer quadrant of the star reduces its brightness during $0.3<\phi<0.6$. However, detailed analysis of these effects awaits an update to the radial-velocity ephemeris, so as to be able to assign reliable absolute phases to photometric features, not just relative ones.

The observed asymmetries are evidently transient effects, as the light curves during other epochs show different deviations from pure ellipsoidal modulation.  In particular, the 2014 December and 2021 November data (Figures~\ref{fig:time-series}a and \ref{fig:time-series}h) have more equal maxima, but show evidence of heating at the L1 point, the minimum at $\phi=0.75$ being higher than the one at $\phi=0.25$.  This is the opposite of what is expected in the absence of heating, when limb darkening and gravity darkening dim the ``nose'' of the companion.  There is some evidence in Figure~\ref{fig:time-series} that the observations of higher flux at $\phi=0.75$ coincide with the flaring episodes, in which case enhanced heating during those epochs might be expected.  However, it is also possible that all of the light curve variations are due to migration of spots.

Such variations appear to be common in those redbacks whose light curves are mainly ellipsoidal, e.g., in PSR~J1723$-$2737, which has been interpreted in terms of differentially rotating starspots \citep{van16}. Very similar behavior is seen in \mspone\ \citep{li14,cho18}.  It has unequal maxima, phase lags, and episodic enhanced flux at $\phi=0.75$, although with a sense of asymmetry different from \obj.  Its peak at $\phi=0.5$ is the higher one, and it is the minimum at $\phi=0.75$ that lags by $\Delta \phi = 0.05$.  Unlike for \obj, the absolute phases of \mspone\ are reliable because it has a radio pulsar ephemeris.  Another example of a double-peaked but highly asymmetric and variable light curve is that of 3FGL J2039.6$-$5618 (PSR J2039$-$5617) \citep{rom15a,sal15,str19,cla21,cor21}.  It can be fitted with one or two spots in addition to pulsar heating.

\subsection{Orbital Eccentricity?\label{sec:disc_b}}

In view of the variable photometric properties of \obj, we revisit the idea that the orbital eccentricity $e=0.04$ detected by \citet{str14} in their radial velocity study of \obj\ is instead an artifact of large star spots or localized pulsar heating, which could bias the velocity toward the center of light and away from center of mass of the star.  This possibility was considered and dismissed by \citet{str14} because, at the time, they had no evidence of pulsar heating or star spots in the light curve as we do now.  On the other hand, even this small eccentricity is unexpected for an old system that should have been circularized by tidal forces.

Consider that when the eccentricity is small, the apparent velocity changes by $\Delta v/v$ of order $e$ around the orbit.  If instead the apparent center of light is biased in certain viewing directions by a fraction $e$ times the orbital separation $a$, a similar effect on the radial velocity will result without the need for any eccentricity. Because \obj\ has the heaviest of the redback companions, its Roche lobe radius is a large fraction of the the orbital separation, $R_L/a=0.33$, which comfortably exceeds $\Delta v/v=0.04$.  If the observed light-curve asymmetries of order 10\% in flux are due to one quadrant of the star being 20\% dimmer than the rest, it would imply that the center of light can be shifted by as much as 10\% of the stellar radius, or $0.033\,a$, which is comparable to that needed to explain the range in velocity without using eccentricity.  A new radial velocity study should be able to test for the persistence of $e$, or instead find that it has changed because it is actually a photometric effect of varying temperature structure on the star.

As an aside, we note that $e=0.04$ would make it essentially impossible to discover pulsations in $\gamma$-rays, a point made by \citet{nie20a} in their analysis of the requirements for such a search of \obj\ given the optical constraints on its orbit.  Discovery of an MSP would be impossible, and even a young pulsar would be difficult.  But if the orbit is circular it should be possible to find the pulsations.

\subsection{Absence of Accretion Disk}

Despite its intense flaring activity, there is no evidence that \obj\ has entered a tMSP state defined by its sub-luminous accretion disk.  The optical emission lines are transient, and the average quiescent X-ray luminosity of $1.2\times10^{32}$~erg~s$^{-1}$, although on the high end for redbacks, is well below the $(6-9)\times 10^{32}$~erg~s$^{-1}$ range of the tMSPs in their disk-passive state, or $(3-5)\times 10^{33}$~erg~s$^{-1}$ in the disk-active state \citep{lin14}.

The UVB colors and orbital variation leave no room for an excess blue continuum from an accretion disk.  The Swift UV filter measurements and upper limits in quiescence are especially restrictive, and can be used to derive an upper limit on the mass transfer rate through a standard sum-of-blackbodies accretion disk \citep{fra02}.  We use the faintest detections and upper limits of $\approx22$~mag in the ${\mathit UVW1}$ as an upper limit on the flux contributed by a disk. We assume that the disk is truncated at the corotation radius of a $1.4 M_{\odot}$ and 2~ms pulsar, $r_{\rm co} = 2.66\times10^{6}$~cm, and is viewed at an inclination angle of $69^{\circ}$.  The resulting limit on the accretion rate is $<1\times10^{14}$~g~s$^{-1}$. This is well below the $\dot m \approx1\times10^{15}$~g~s$^{-1}$ rate estimated by \citet{pap15} for the original tMSPs PSR J1023+0038 and XSS J12270$-$4859 using similar assumptions.

\subsection{Characteristics and Origin of Flares\label{sec:disc_c}}

\begin{figure}
\vspace{-0.3in}
\centerline{
\includegraphics[angle=0.,width=0.95\linewidth]{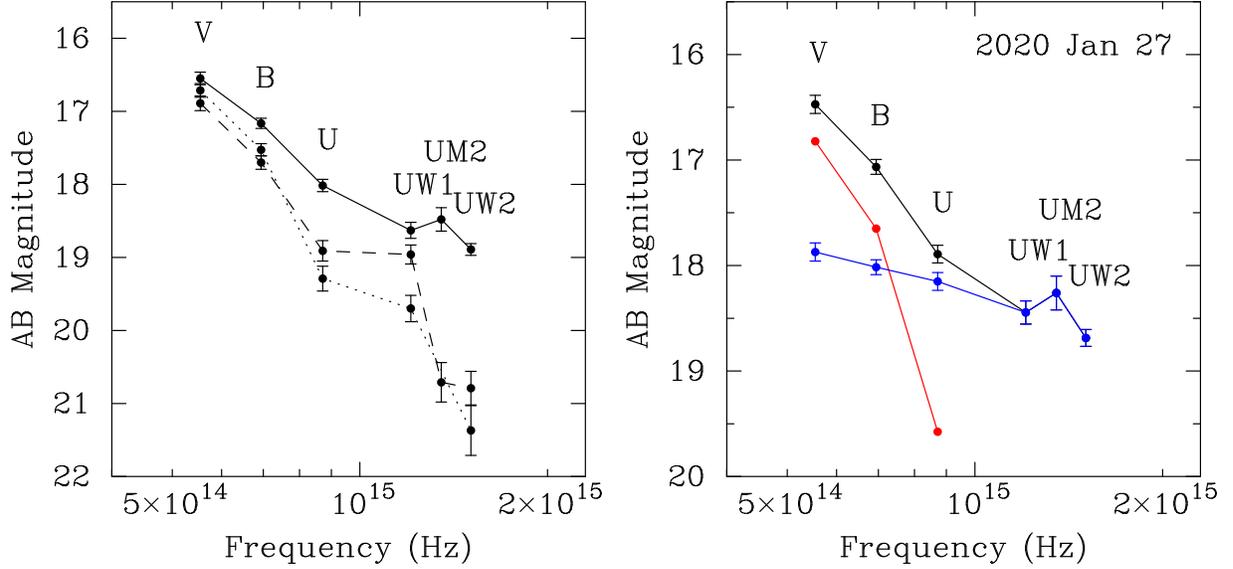}
}
\vspace{-1.9in}
  \caption{
  Left:  UVOT magnitudes of the flare state of 2020 January 27 (Figure~\ref{fig:swift_1}).  Lines connect the points from the same GTI: first (solid), second (dashed), and third (dotted). Right: Spectral energy distribution of the first GTI, corrected for interstellar extinction (black points). Red points are quiescent $U, B$, and $V$ magnitudes corresponding to the same orbital phases.  These are subtracted to isolate the spectrum of the flare (blue points), which can be fitted with a power-law, $f_{\nu}=\nu^{-0.71\pm0.09}$.
  }
\label{fig:uvot_flare}
\end{figure}

One of the flares caught by \swift\ has sufficient coverage in the UVOT filters to quantify its spectral energy distribution.  The data on 2020 January 27 (Figure~\ref{fig:swift_1}) has observations in all six filters near the maximum of the X-ray flux, as well as during the decay.   In Figure~\ref{fig:uvot_flare} (left) we graph the magnitudes during each of the three GTIs, even though they are not strictly simultaneous.  

In order to isolate the spectral energy distribution (SED) of the flare from that of the star, we have to subtract the contribution of the photosphere to each filter, which can be done by using magnitudes from the ellipsoidal light curves in $U,B,V$ at the appropriate orbital phases.  The stellar contribution in the UV bands is negligible, as the quiescent state is not detected in these filters.  We corrected all magnitudes for interstellar extinction assuming $E(B-V)=0.025$, and used extinction coefficients for the UVOT filters appropriate for low extinction and hot stars (given the blue colors of the flare) from \citet{sie14}.  These are $A_{\lambda}/E(B-V) = 3.10, 3.99, 4.96, 7.40, 8.79$, and 8.71 for ${\mathit V, B, U, UVW1, UVM2}$, and ${\mathit UVW2}$, respectively.

Figure~\ref{fig:uvot_flare} (right) shows, for the peak of the flare, the stellar contribution in red, and the net flare spectrum in blue.  The flare spectrum is fitted well by a power law of energy index $\alpha=0.71\pm0.09$, where $f_{\nu}\propto\nu^{-\alpha}$. Although decaying flux is still detected in the UV during the subsequent GTIs, it is not possible to reliably characterize the SED in this phase of the flare, as the $U,B,V$ contribution reverts to being dominated by the star.

The UV and X-ray spectra at the peak of the flare are each fitted by a power law of energy index 0.7, although they do not connect precisely.  The UV flare spectrum lies a factor of $\approx 3$ above an extrapolation of the X-ray spectrum.  Alternatively, to connect the UV and X-ray fluxes at $10^{15}$~Hz and 1~keV, respectively, a spectral index $\alpha_{\rm ox} = 0.91$ would be required.  This value is marginally consistent with the fitted X-ray index of the flare state in Table~\ref{tab:xspec}. If such an interpolation is valid, then the integral of the optical through X-ray flux at the peak of the flare of 2020 January 27 is $\approx2.1\times10^{-11}$ erg~cm$^{-2}$~s$^{-1}$, corresponding to a luminosity of $\approx1.3\times10^{34}$ erg~s$^{-1}$. 

This flare is of different character from flare seen in an optical spectrum of the black widow PSR~J1311$-$3430 \citep{rom15b}, which shows a transient spectrum of a He-rich, $39,000$~K stellar atmosphere, as well as \ion{He}{1} emission lines.  The authors attributed this spectrum to a heating source well below the photosphere, e.g., magnetic reconnection, but triggered externally by interaction with the pulsar wind.  In contrast, the optical spectrum of \obj\ taken during the peculiar, rapidly flaring episode on 2020 February~7 (Figure~\ref{fig:spectra}), while having enhanced emission lines, does not show significant change in its continuum shape (apart from having weaker absorption lines).  This supports the evidence from the UVOT spectral energy distribution that the optical flaring in \obj\ is nonthermal, coming from above the photosphere, or at least not affecting a significant fraction of the stellar surface as it does in PSR~J1311$-$3430, a much smaller star, with $R\approx0.07\,R_{\odot}$ \citep{rom15b}.  However, the companion star in \obj\ is much more luminous; its quiescent magnitudes given in Section~\ref{sec:disc_a} correspond to almost exactly $1\,L_{\odot}$, while the peak UV and X-ray luminosity of a flare is $\sim3\,L_{\odot}$.  So we cannot rule out that localized photospheric heating makes some contribution to the UV spectrum of the 2020 January 27 flare in \obj, which exceeds somewhat the extrapolation of its best-fit X-ray spectrum.

We can ask how the flares observed only in MDM $R$-band photometry compare with those observed only by Swift at shorter optical and UV wavelengths. A illuminating example is the end of the brightest flare on 2020 January~24 (Figure~\ref{fig:time-series}b), which occurred just 3 days before the bright January~27 flare seen in Swift (Figure~\ref{fig:swift_1}).  The light curves of these are similar, having rise and fall times of $\sim5-10$ minutes in the optical or X-ray.  The amplitude of the strongest $R$-band flare is 0.35~mag above the quiescent state (this event was also seen by ZTF), while the Swift flare rose by almost exactly that amount in the $V$ band (Figure~\ref{fig:uvot_flare}).  So it is likely that these are of similar broad-band luminosity, and representative of other flares observed only from the ground.

Of principal interest is whether the flares are ultimately powered by the rotational energy of the pulsar, and if they require the release of magnetic energy, e.g., via reconnection of striped magnetic field compressed in the shocked pulsar wind \citep{sir11,cor22}.  Alternatively, the reservoir of magnetic energy in the companion may be greater, renewable, and capable of channeling the pulsar wind \citep{san17}.  There is some evidence for magnetic reconnection in the X-ray spectra. The flares in redbacks \mspthree\ and \msptwo\ \citep{hal17a,cho18} and the black widow PSR~J1311$-$3430 \citep{an17} have flatter spectra than their quiescent emission, with flare photon index $1.2<\Gamma<1.5$.  While the particle energy distribution needed to emit synchrotron with $\Gamma<1.5$ is difficult to achieve by diffusive shock acceleration, it has been produced in simulations of magnetic reconnection cited above.
It is surprising therefore that the flare states in \obj\ have $\Gamma=1.7$, steeper than that of its intermediate and quiescent states, which have $1.3<\Gamma<1.5$, more typical of other spiders.  Considering the extreme properties of this system, it may be an indication that there is an additional emission process contributing to the flare spectrum.

A peak flare luminosity of $\sim 10^{34}$~erg~s$^{-1}$ is comparable to the spin-down power of redbacks, which fall in the range $10^{34}-10^{35}$~erg~s$^{-1}$ \citep{str19}, while only a fraction of the pulsar wind is intercepted by the companion or its stellar wind.  In the rotation-powered scenario, this already implies either a very energetic pulsar, or a large cross section for the target.  Even the large estimated Roche-lobe radius, $R_L = 0.33a$ for the relatively massive companion star, which implies that it subtends a fraction $\approx(R_L/2a)^2 = 0.027$ of the solid angle seen by the pulsar, would require \obj\ to be the most energetic redback pulsar unless the spin-down energy is highly beamed in the orbital plane.

Alternatively, if the companion wind is very extensive, being wrapped around the pulsar or even completely enclosing it, then this requirement is relaxed.  Such ``hidden MSPs'' that could be detectable as high-energy sources but invisible in radio were predicted by \citet{tav91} and \citet{tav93}. Interestingly, the same considerations apply to the ``huntsman'' system 3FGL J1417.5$-$4402 (PSR J1417$-$4402), a 2.66 ms pulsar with a giant companion \citep{str16,cam16,swi18}.  To power its large $\gamma$-ray luminosity, the pulsar spin-down power (which has not yet been measured) must be one of the highest among MSPs, or else the $\gamma$-rays are mostly produced in an extensive wind shock rather than being pulsed emission from the magnetosphere. In any case, an extensive wind is actually inferred from the fact that radio pulsations are infrequently detected from PSR J1417$-$4402.

If there are any short- or long-term trends in the rate of occurrence of flares in \obj, it is not clear due to its spotty history of observation.  First, there does not seem to be a preference for certain orbital phases. Either the location of flaring sites is not restricted, or the emission is not highly beamed.   In other redbacks, \msptwo\ and \fgl, it does appear that states of quiescence and flaring cycle on timescales of weeks or months \citep{cho18}.  There may be some indication of such a distribution in the MDM light curves of \obj, but we do not yet have enough dense coverage to establish it.  Similarly, the occurrence of two of the four flares in the ASAS-SN data within 17 days of each other is suggestive, but not highly significant.

On a time scale years, it is hard to judge because the CRTS (2006--2013), ASAS-SN (2014--2018), and ZTF (2018--2021) have different sampling patterns from our dense MDM time series, which began only in 2020.  Only continued monitoring of \obj\ can determine if its flaring rate or intensity is growing in the long term, which is of interest in case flaring is related to the stage of evolution of a redback relative to the sub-luminous disk state of the tMPSs.  So far, there is no evidence from tMSPs that flaring preferentially precedes or follows a transition. 

\section{Conclusions\label{sec:conc}}

We discovered luminous flaring from the putative redback \obj\ during ground-based time-series monitoring in 2020 January, but it can also be seen occasionally in sky-survey data going back to 2007.  The duty cycle of the flares is high enough that triggered \swift\ monitoring was able was able to detect several similar flares and intermediate level activity over a period of 100~days.  After that, sparser optical and X-ray monitoring continued to show intermittent activity through the end of 2021, suggesting that this is regular behavior which is accessible for more detailed study.

An early result of this program was the discovery of intermittent optical emission lines and their correlation with the flaring state, an association that had only been suspected previously for redbacks based on non-contemporaneous observations of optical or X-ray flares and optical spectroscopy.  Flares detected simultaneously in the X-ray and optical/UV with \swift\ appear to have power-law spectra and peak broad-band luminosities of $\sim10^{34}$~erg~s$^{-1}$, the largest among redbacks.  Although such luminosity overlaps with the subluminous disk states of the tMSPs, there is no evidence of a disk in the quiescent optical, UV or X-ray emission.

The X-ray spectra in the quiescent and intermediate states of \obj\ are typical of spiders, and flat enough ($\Gamma\approx1.4$) to suggest that magnetic reconnection is occurring.   However, the flare spectrum with $\Gamma\approx1.7$ is steeper than those of flares in other spiders, which suggests it might have a contribution from an additional emission mechanism.  \obj\ has the most massive companion star of the known and candidate redback systems, which may provide ingredients needed to trigger luminous flares.  These could include a well-developed stellar wind as a target for shocks, and perhaps significant magnetic field energy of its own. 

Apart from the flaring states, the optical light curve shapes, which deviate from pure ellipsoidal modulation, are suggestive of large starspots or variable heating by the pulsar.  These effects are common in redbacks.  Because of the large size of the companion in \obj\ relative to the diameter of the orbit, such asymmetries should be considered as a possible distorting effect on the radial velocity curve. The spin-down power of the pulsar is an important unknown factor in the energy budget of the flares and any persistent heating of the companion's photosphere.  Detection of radio or $\gamma$-ray pulsations would add this missing piece.  In addition, radio pulsations could be used to diagnose the angular extent, density, and variability of the stellar wind, unless the wind is too perfect a shroud and absorbs the radio pulses completely.

\begin{acknowledgements}

 We thank the anonymous referee for several helpful suggestions. This work is based on observations obtained at the MDM Observatory, operated by Dartmouth College, Columbia University, The Ohio State University, Ohio University, and the University of Michigan.  Support for this work was provided in part by NASA grants 80NSSC20K1108 and 80NSSC21K1380 awarded through the Swift Guest Investigator Program. This research has made use of data and software provided by the High Energy Astrophysics Science Archive Research Center (HEASARC), which is a service of the Astrophysics Science Division at NASA/GSFC and the High Energy Astrophysics Division of the Smithsonian  Astrophysical Observatory. We acknowledge use of NASA's Astrophysics Data System (ADS) Bibliographic Services and the arXiv.

\end{acknowledgements}

\end{document}